\RequirePackage{fix-cm}
\documentclass[smallextended]{svjour3}       
\smartqed  
\usepackage{graphicx}
\usepackage{caption}

\usepackage{color}
\usepackage{natbib}

\usepackage[caption=false]{subfig}
\usepackage[lmargin=3.0cm, rmargin=3.0cm,tmargin=2.50cm,bmargin=2.50cm]{geometry}

\usepackage[labelfont=bf]{caption}
\newcommand\Mo{M$_{\odot}$}
\newcommand{\aap}{A \& A}
\newcommand{\pasp}{PASP}
\newcommand{\mnras}{MNRAS}
\newcommand{\apj}{APJ}

\newcommand{\araa}{ARA\&A}
\newcommand{\qjras}{QJRAS}
\newcommand{\aplett}{Astrophys. Lett.}
\newcommand\micron{\ensuremath{\mbox{$\mu$}\mbox{m}}}

 \journalname{Experimental Astronomy}
\begin{document}

\title{An automated approach for photometry and dust mass calculation of the Crab nebula
}

\titlerunning{Calculating the dust mass in Crab nebula}        

\author{Cyrine Nehm\'e         \and
        Sarkis Kassounian \and Marc Sauvage 
}


\institute{Cyrine Nehm\'e   \at
              Department of Physics \& Astronomy, Notre Dame University - Louaize, Lebanon \\
              AIM, CEA, CNRS, Universit\'e Paris-Saclay,Universit\'e Paris Diderot, Sorbonne Paris Cit\'e, F-91191 Gif-sur-Yvette, France\\
              \email{cnheme@ndu.edu.lb}           
           \and
           Sarkis Kassounian \at
              Department of Physics \& Astronomy, Notre Dame University - Louaize, Lebanon\\
           \and
           Marc Sauvage \at
           AIM, CEA, CNRS, Universit\'e Paris-Saclay,Universit\'e Paris Diderot, Sorbonne Paris Cit\'e, F-91191 Gif-sur-Yvette, France  
}

\date{Received: date / Accepted: date}

\maketitle

\begin{abstract}
Ample evidence exists regarding supernovae being a major contributor to interstellar dust. In this work, the deepest far-infrared observations of the Crab Nebula are used to {\bf revisit the estimation of} the dust mass present in this supernova remnant. Images in filters between 70 and 500 \micron\ taken
by the PACS and SPIRE instruments on-board of the Herschel Space Observatory are used. {\bf With} an automated approach we constructed the 
spectral energy distribution of the Crab nebula to recover the dust mass. This approach makes use of several image processing techniques (thresholding, morphological processes, contouring,  etc..) to objectively separate the nebula from its surrounding background.
After subtracting the non-thermal synchrotron component from the integrated fluxes, the spectral energy distribution is found to be best fitted using a single modified blackbody of temperature $T=42.06\pm1.14$\,K and a dust mass of 
$M_{d}=0.056\pm0.037$\,\Mo. 
In this paper, we show the importance of the photometric analysis and spectral energy distribution construction in the inference of the dust mass of the Crab nebula.  

\end{abstract}

\section{Introduction}
\label{intro}
Stars in their Asymptotic Giant Branch (AGB) phase have been recognized as important sources of dust \citep[with typical yields of $0.01-0.05$\,\Mo,][]{Helling2006,Andersen2007,Gail2009,Schneider2014,Dell'Agl2015}. But the amount of dust produced and ejected {\bf into} the interstellar medium by these stars is not enough to compensate for the known destruction rates \citep{Zhukovska2008}. Stellar dust injection by itself also cannot explain the amount of dust in high-redshift objects where star formation has not been going on for large periods of time for a significant AGB star population to exist. Hence the dust mass calculated from observations must have other sources that can compensate this discrepancy.\\

In the literature, evidence has been presented which favors the concept of supernovae being a major contributor for dust formation \citep{Krafton2016}. This is supported by nucleation models \citep{Silvia2010,Marassi2015}, where dust is formed in dense regions during the first phases of supernovae explosions and allows explaining the large quantities of dust found in galaxies \citep{Dwek2010}. Theoretical models predict that supernovae can be efficient dust producers ($0.1-1.0$\,\Mo), potentially accounting for most of the dust production in the early Universe \citep{DeLooze2017}. Observational evidence for this dust production efficiency is however currently limited to only few supernova remnants (SNR).\\

To revisit this issue, the Crab nebula presents many advantages: as one of the most commonly observed celestial objects, it possesses a wealth of observational data and at a relatively young age of the remnant, the mass of swept-up material is small compared to the mass in the supernova ejecta, which makes it possible to separate supernova dust from any swept-up circumstellar material. It also provides the cleanest view along the line of sight compared to other observed SNRs. It is a pulsar wind nebula (PWN) that is sweeping up supernova ejecta material observed in the form of bright optical filaments \citep{Gaensler2006}. Unlike other remnants, there is almost no dusty galactic material in front of, or behind the Crab Nebula, so the images are less effected by scattering or extinction {\bf (though we will see that the environment is far from innocuous)}. {\em Spitzer} infrared observations at short wavelengths allowed the discovery of warm dust in the filaments but it missed a potentially more massive cool dust component. {\em Herschel}, observing at longer wavelengths, is better suited for that purpose, allowing astronomers to measure the total amount of dust for the first time.\\ 

\label{literature} 
\cite{Trimble1977} reported the first evidence for the existence of dust in the Crab Nebula. This emerged from the existence of infrared (IR) excess emission in the synchrotron spectrum. In further studies, discrepancies were noted in the dust masses based on different theoretical modelling assumptions. For example, in \cite{gomez2012}, the dust component observed along the ionized filament structure in spatially coincidence with the location of ejected material was modelled with modified black-bodies (MBB). A fit with a single MBB model lead to the derivation of 0.08\,\Mo\ to 0.14\,\Mo\ depending on the choice for dust composition (respectively amorphous carbon or silicate). This model however provided a poor fit to the entire spectral energy distribution (SED). A more adequate fit was obtained with two temperature components, leading to masses of 0.11\,\Mo\ for amorphous carbon grains to 0.24\,\Mo\ for standard silicates (we note that, in that paper, the 24\,\micron\ flux was included in the fitting process, which is debatable for a model assuming the emission is coming from a black-body). In \cite{tea2013}, the total dust mass derived from the best-fit models to the observed IR spectrum ranged between 0.019\,\Mo\ and 0.13\,\Mo. In their work, the models compared three different dust compositions, each characterized by a power-law distribution in grain radii whose index was allowed to vary. In \cite{Owen2015} several models were applied and compared with a variety of grain dust compositions and grain size power law spectral indexes ranging from 2.7 to 3.5, resulting in dust mass values ranging between 0.1 to 1.0\,\Mo.\\ 

It is thus clear that the dust content of the Crab nebula is still not well constrained in the  literature, hence we revisit in this work the estimation of the dust content in the nebula using improved photometric images and image analysis techniques, and attempting at a proper quantification of all the uncertainties associated to the process. 
We have applied an automated image analysis technique to optimize the extraction of the integrated object flux density at each wavelength. This approach increases the accuracy of the flux densities by excluding as much as possible the background flux contamination by edge detection processes. This paper is arranged as follows: in Section~\ref{sec:obs}, we will describe the data and processing used to produce the improved images, in Section~\ref{sec:photom} image analysis and flux calculation will be explained, Section~\ref{sec:dust} will present the model used for the dust mass calculation, Section~\ref{sec:irex} will describe the calculations of the mass and temperature of dust, and finally in Section~\ref{sec:concl} we discuss our result and conclude this study.\\

\section{Observations and Data} 
\label{sec:obs}

Images used for the calculation of the flux densities of the Crab nebula were taken by the \textit{Herschel Space Observatory} \citep{herschel2010}. The Crab nebula was observed between September 2009 and September 2010. The PACS \citep{PACS2010} and SPIRE \citep{SPIRE2010} instruments performed photometry at 70, 100, 160, 250, 350 and 500\,\micron, as part both of a calibration program for the PACS observing modes, and of a {\em Principal Investigator} observing program \citep[see][]{gomez2012}. A summary of the observations is listed in Table \ref{table: observations}.\\

\begin{table}[h]
\centering
\begin{tabular}{c c c c c}
\hline
\hline\\ \vspace*{0.2cm}
Obs. ID & Obs. Date & Obs. Duration (s) & Instrument & Filter set $\lambda$ (\micron) \\

\hline
\hline\\

1342183905 & 2009-09-15 & 2221 & PACS & (70, 160) \\
1342183906 & 2009-09-15 & 2221 & PACS & (70, 160) \\
1342183907 & 2009-09-15 & 2221 & PACS & (70, 160) \\
1342183908 & 2009-09-15 & 2221 & PACS & (70, 160) \\
1342183909 & 2009-09-15 & 2221 & PACS & (100, 160) \\
1342183910 & 2009-09-15 & 2221 & PACS & (100, 160) \\
1342183911 & 2009-09-15 & 2221 & PACS & (100, 160) \\
1342183912 & 2009-09-15 & 2221 & PACS & (100, 160) \\
1342191181 & 2010-02-25 & 4555 & SPIRE & (250, 350, 500) \\
1342204441 & 2010-09-13 & 1671 & PACS & (70, 160) \\
1342204442 & 2010-09-13 & 1671 & PACS & (100, 160) \\
1342204443 & 2010-09-13 & 1671 & PACS & (70, 160) \\

\vspace*{0.2cm}
1342204444 & 2010-09-13 & 1671 & PACS & (100, 160) \\ 

\hline
\hline\\

\end{tabular}

\caption{Photometric observations of the Crab Nebula using the \textit{Herschel Space Observatory}.}
\label{table: observations}
\end{table}

The PACS photometry data were obtained in scan-map mode with speed 20 arcsec/s. In order to obtain images in all three filters of the PACS photometer, observations are done by pairs, one using the 70 and 160\,\micron\ filters, and one using the 100 and 160\,\micron\ filters. Therefore the 160\,\micron\ image has twice the exposure time as the other two channels. In the present work we have for the first time included data that was obtained on the Crab nebula during the testing and qualification of PACS observing modes. As these data were obtained in an instrumental set-up which is identical to the operational one, they can be directly combined with the already published data, thus nearly doubling the depth of the resulting maps (at 70, 100, and 160\,\micron). The resulting PACS maps have a size of $25'\times25'$, {\bf although the full depth of the maps is only achieved on a smaller field of view as not all observations mapped the same area}. The SPIRE maps were obtained using the Large Map mode with a scan length of 30 arcmin at a speed of 30 arcsec/s, to produce a map of $32'\times32'$.

\begin{figure}[h]
\centering
\includegraphics[scale=0.5]{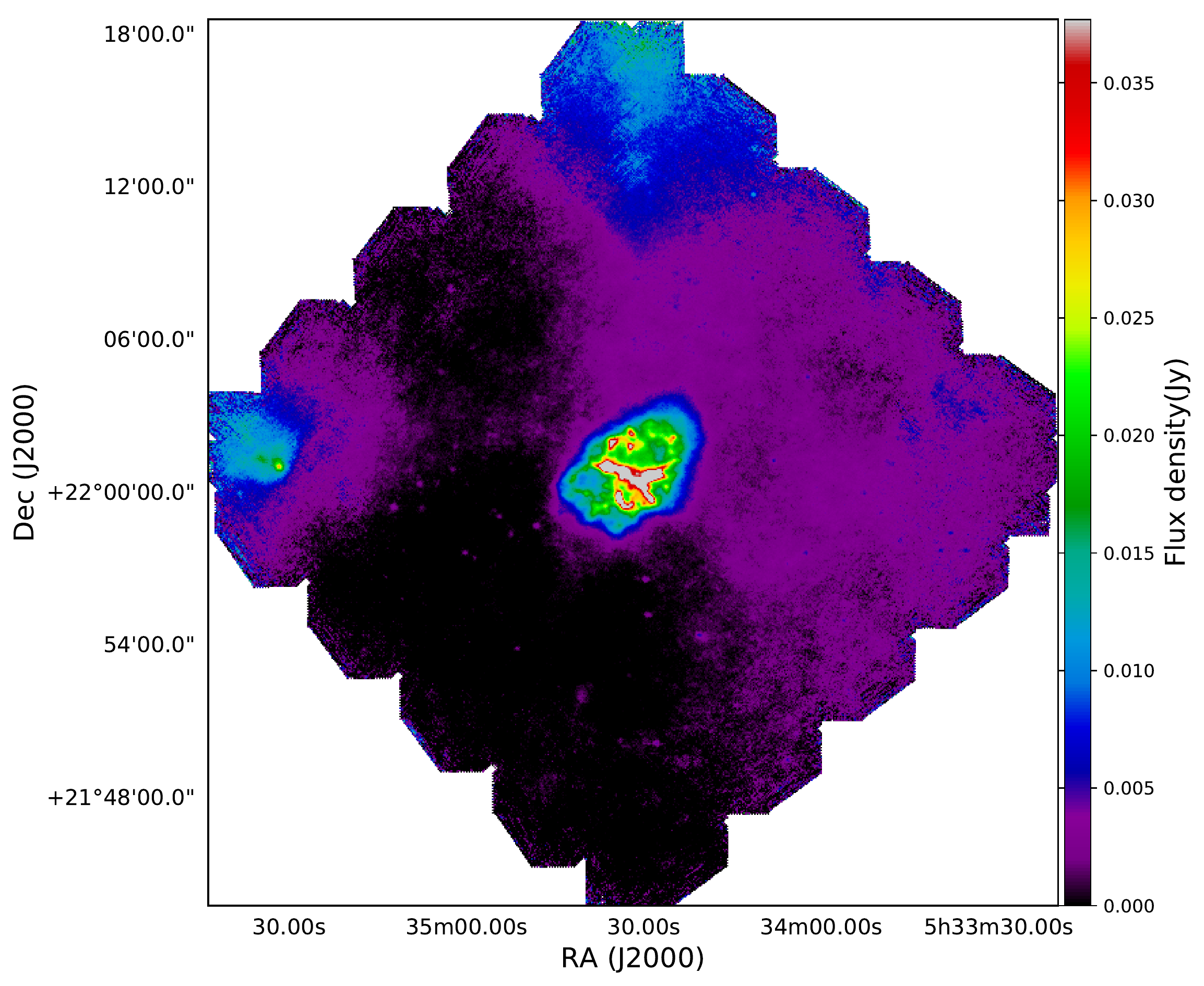}
\caption{The full Crab nebula field of view (FOV) at 160\,\micron. On this image the complex background is quite visible in the NW side of the image, as well as an unknown, unrelated, object near (+22$^\circ$00',5h36m) residing besides the Crab nebula. This object is quite prominent at long wavelengths, and essentially undetected shortward of 160\,\micron. {\bf The brighter area at the top of the image is the combination of actual structure and an artefact of the data processing.}
\label{fig:obj}}
\end{figure}

Figure~\ref{fig:obj} displays the resulting map at 160\,\micron\ as it is quite representative of what we observed in the field. In almost all our images, there exists two objects: the Crab nebula and an unknown celestial object to its {\bf eastern} side, as well as an extended background/foreground emission to the NW that reaches to the nebula itself. This unknown object has a significant IR flux especially in the long wavelengths bands and will need to be masked out when we will try to fit the structure of the extended background.\\

The PACS data were first reduced using the Herschel Interactive Processing Environment \citep[HIPE, see][]{Ott2010}. Then the Unimap software \citep{Piazzo2015}, which is a generalized least-squares map maker for Herschel data, was used to eliminate low frequency noise in the maps and combine all PACS observations in a single map per photometer band. For SPIRE, the standard pipeline HIPE version 14.0.0 was used \citep{SPIRE2010}. \\ 

\begin{figure*}[htp]
\center
   {\subfloat[$\lambda$=70\,\micron\ \& pix size=1.6''
]{  
      \includegraphics[width=.45\textwidth]{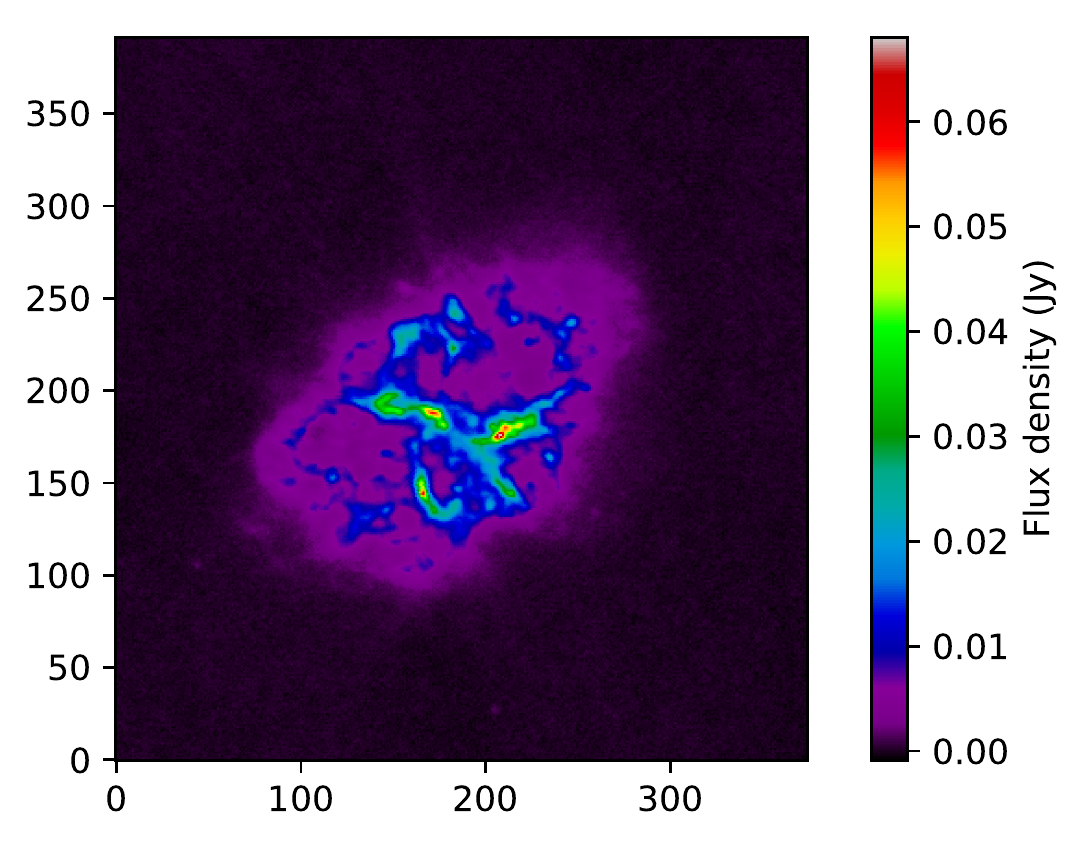}}
~
 \subfloat[$\lambda$=100\,\micron\ \& pix size=1.6''
 ]{  
      \includegraphics[width=.45\textwidth]{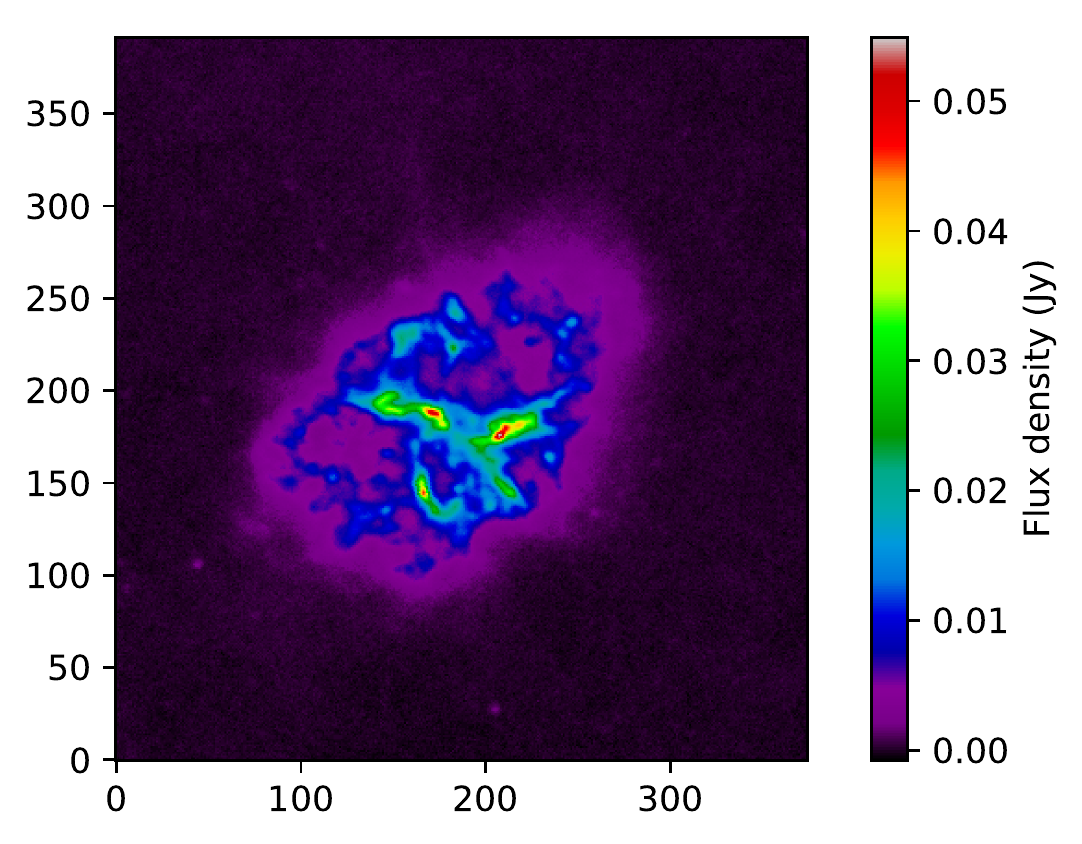}}

}

\center
   {\subfloat[$\lambda$=160\,\micron\ \& pix size=3.2''
 ]{  
      \includegraphics[width=.45\textwidth]{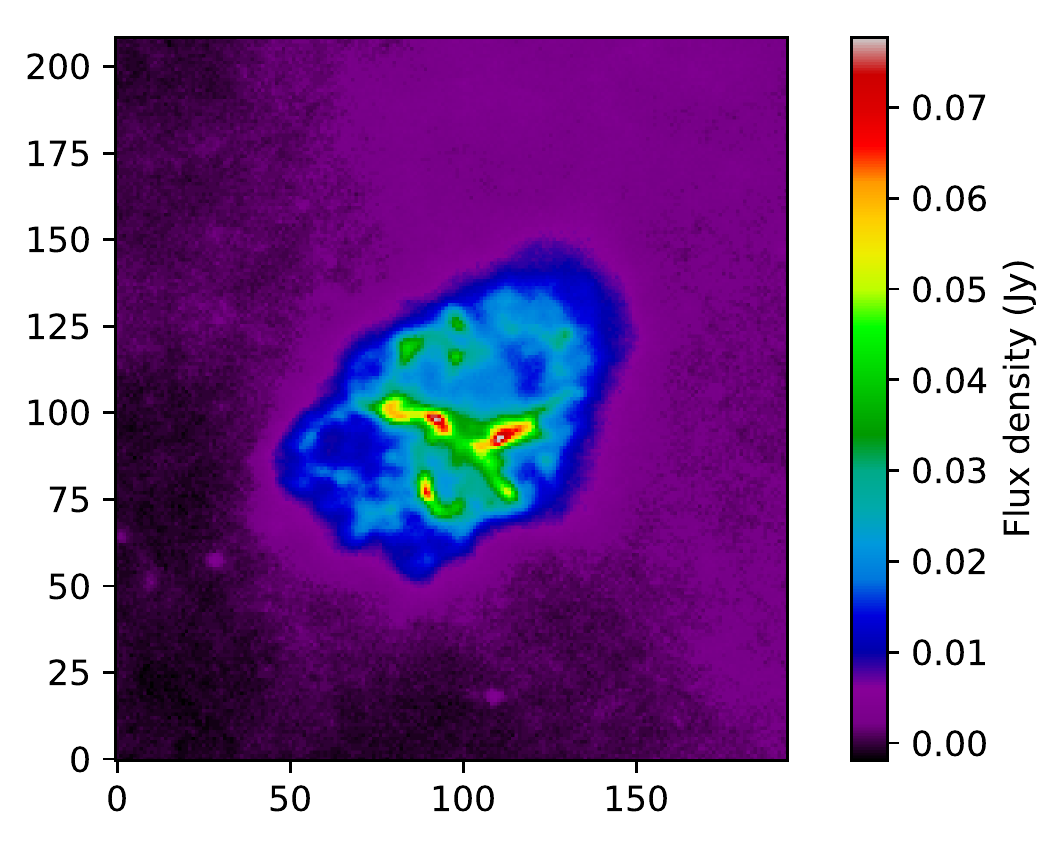}}
~
   \subfloat[$\lambda$=250\,\micron\ \& pix size=6.0''
 ]{  
      \includegraphics[width=.45\textwidth]{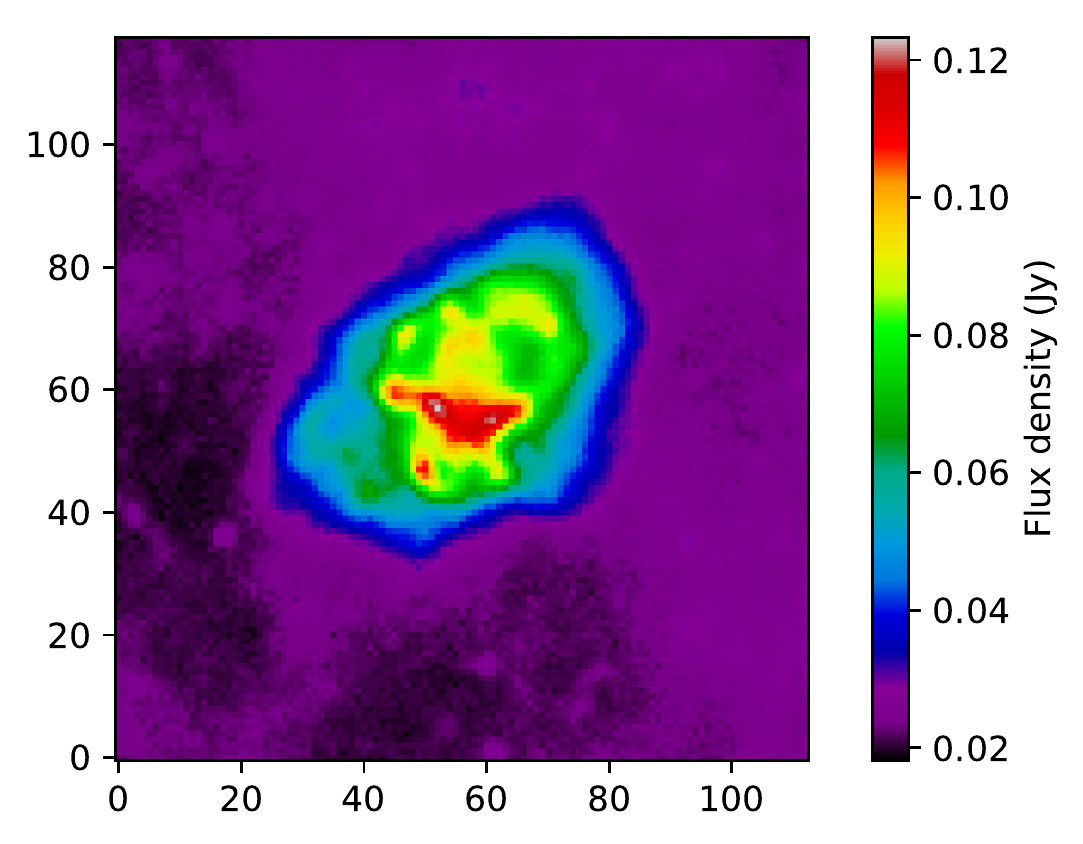}}
   }

 \center
   {\subfloat[$\lambda$=350\,\micron\ \& pix size=10.0''
 ]{  
      \includegraphics[width=.45\textwidth]{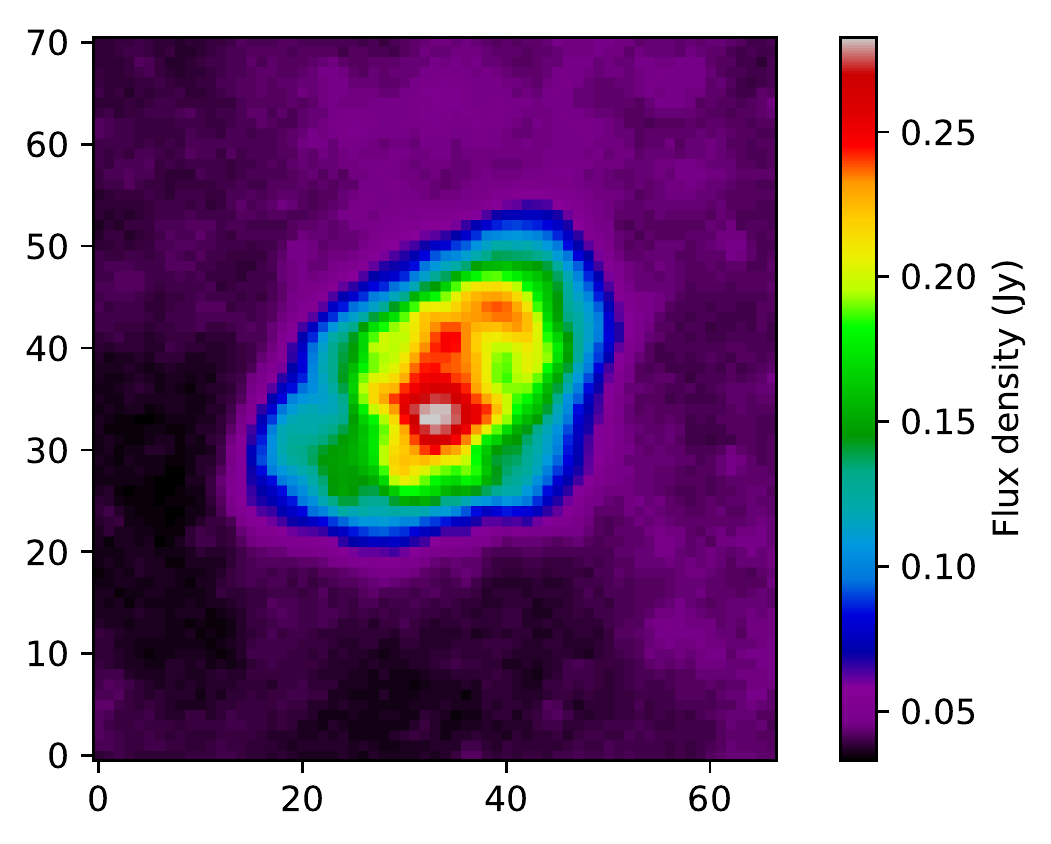}}
~
   \subfloat[$\lambda$=500\,\micron\ \& pix size=14.0''
 ]{  
      \includegraphics[width=.45\textwidth]{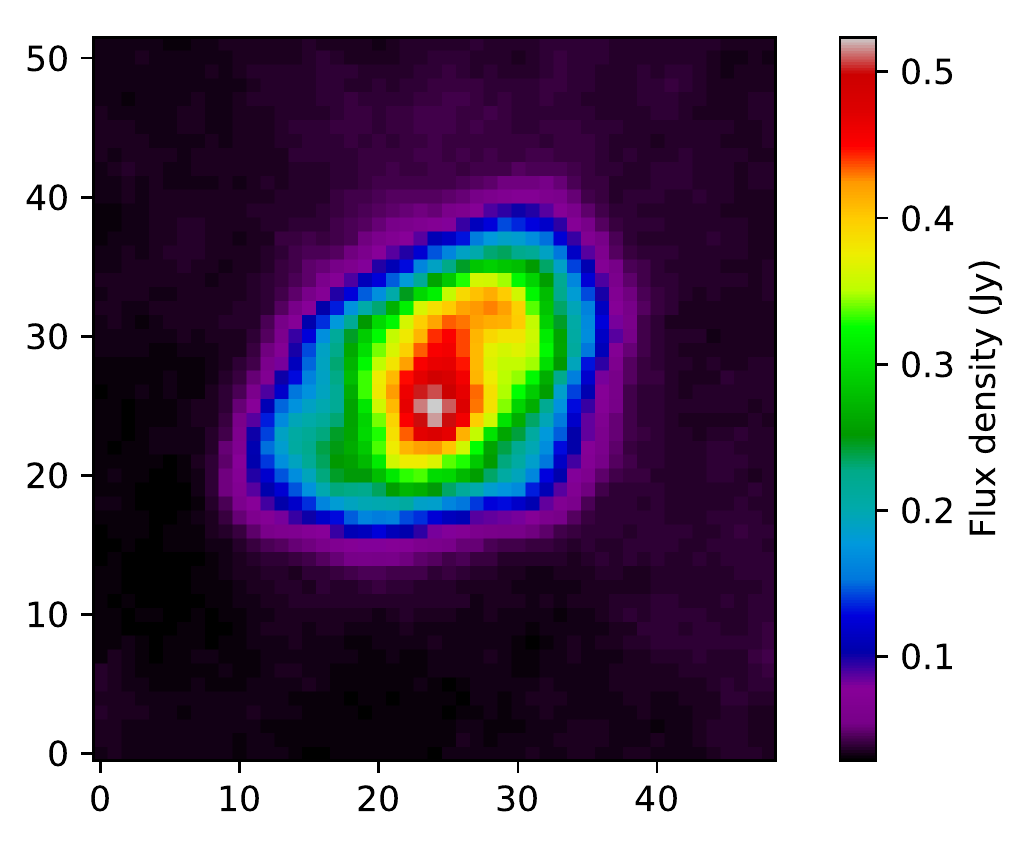}}
   }

\caption{This figure presents the central areas of maps associated to  PACS (70, 100, and 160\,\micron) and SPIRE (250, 350, and 500\,\micron),
zooming in on the Crab nebula. The color codes the flux density (Jy) per pixel. The x and the y axis are pixel coordinates. In each subcaption the wavelength and associated pixel size are mentioned.  }
\label{fig: zooms}
\end{figure*}

\section{Photometric Image analysis}
\label{sec:photom}

\textbf{The main challenge regarding the photometry of extended objects in the far infrared (FIR) is the segmentation and the separation of the object of interest in foreground from that of background emission (note that we use these terms, foreground and background, in a generic way as some of the structure unrelated to the Crab Nebula could be in front of it). A proper estimation and removal of non-intrinsic contributions to the object of interest is very important. }The Crab nebula observations are no exception: Fig.\,\ref{fig: zooms} shows significant emission in the immediate vicinity of the nebula (this is clearly visible from 160 to 350\,\micron\ but this emission is {\bf detected} at {\bf nearly} all wavelength).\\

To calculate the integrated flux densities of the Crab nebula at each wavelength, the standard approach is applying aperture photometry with an aperture based on the shape of the object under study (such as circle or an ellipse) but for extended objects this means including higher number of background pixels which can create a substantial error by accumulation. Therefore it is necessary to detect the edges of the Crab nebula morphologically and as objectively as possible. The iterative procedure that we present here aims at detecting the outer edges of the Crab nebula, through the adaptation of image analysis techniques for edge detection. It also aims at increasing the accuracy of the integrated flux density by {\bf reducing} the inclusion of background pixels.\\

\textbf{Independently for each photometric band, our adopted edge detection technique is based on computing a threshold flux density value that will classify the pixels of the photometric image as either belonging to the object of interest or background. In other words, the algorithm searches for the transition between pixels dominated by the foreground object (Pulsar synchrotron and dust IR emission) and that of the background (Galactic sources) by searching for the location in the image where variations in pixel brightness are tracing the wings of the PSF applied to the extended object.} The successive steps of the method we have retained are detailed in the following subsections.

\subsection{Initial choice of an image thresholding level}

An initial test was done using a classical image segmentation technique called Otsu's method \citep{Otsu}. It utilizes the statistical analysis of the pixel intensity histogram of images to classify the pixels into two categories or distributions(background and foreground). Otsu's method is an algorithm that tries to find a separation criteria or threshold value for the flux densities of the image assuming that it has a bimodal nature. Using this assumption, it derives a flux threshold value that maximizes the intraclass variance and minimizes the interclass variance of the data to generate a clear separation of the image in two sets of areas. Once this optimal threshold is determined, a mask is built from the map where pixels above the threshold are set to 1 (foreground) and pixels below are set to 0 (background), see Fig.\,\ref{fig:otsu} (a) and (b). Most astronomical images do not have a bimodal distribution, however, and situations of uni or multi-modality can exist. {\bf The histograms of Herschel maps have a common nature (see Fig.~\ref{fig:otsu}a): the low background flux density values dominate with the highest occurrence rates, whereas the higher flux densities associated to the filaments or the pulsar synchrotron provide a long tail of less populated bins at high intensity values.}\\

\textbf{For this  reason, the region initially selected as foreground using Otsu's threshold ($\theta_{otsu}$) is not optimal for deriving the photometry of the Crab Nebula: it is creating voids in the inner parts of the nebula and rejecting outside of the mask fainter emission regions that still obviously belong to the Nebula (see Fig.\,\ref{fig:otsu}). Thus $\theta_{otsu}$ is rather an  estimation for the initial guess needed for the adopted iterative optimal threshold determination method that we will describe in section~\ref{sec:find theta_op}}.

\subsection{ Opening and isophotal contouring}

\begin{figure*}[b!]
\center
\hspace*{-1cm}
   {\subfloat[Histogram of flux density values. The bins go up to 0.06 Jy/pixel, but the x-axis has been cut since they are too scarcely populated to be seen on this graph.]{  
      \includegraphics[width=.55\textwidth]{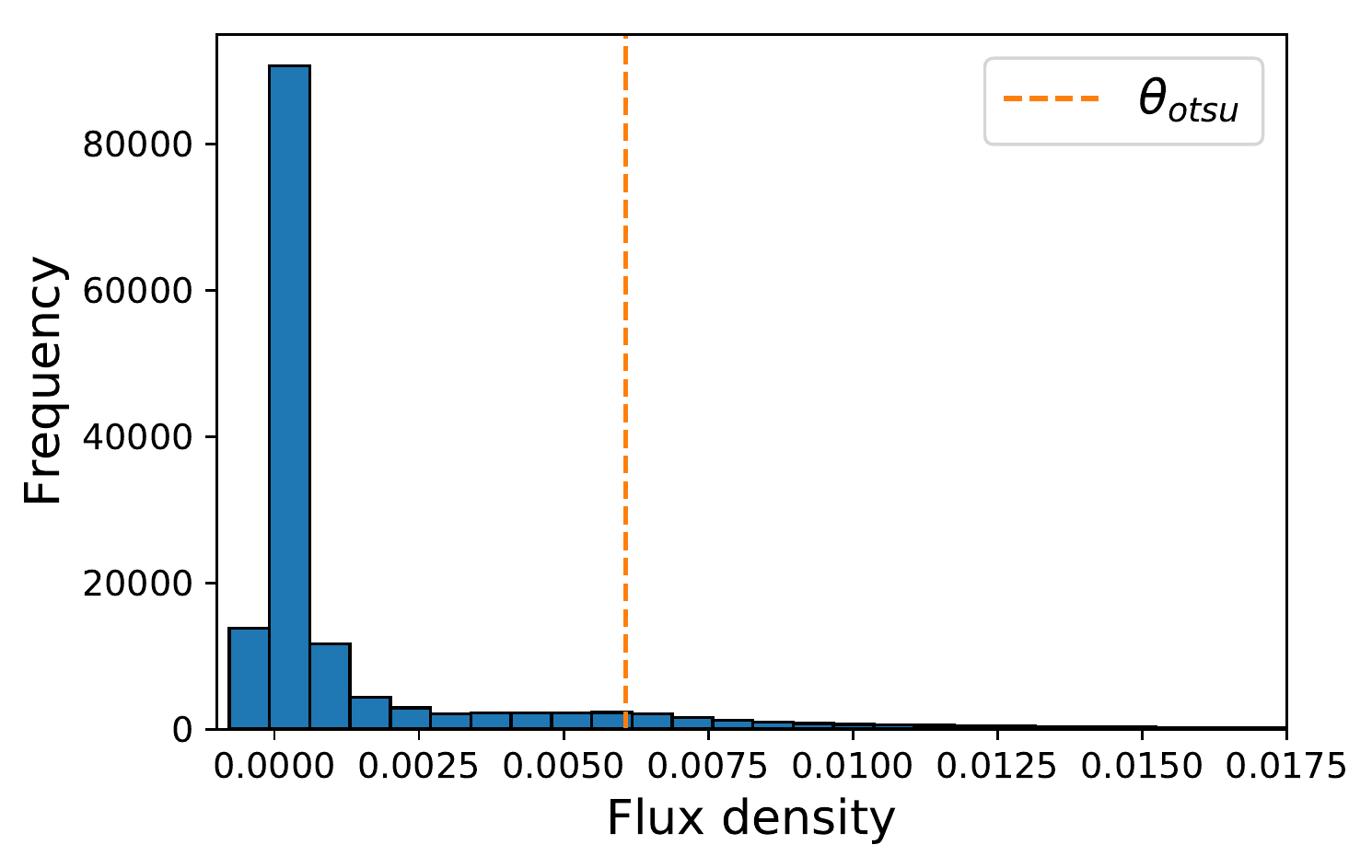}}
~
 \subfloat[ Generated mask based on Otsu's threshold ]{  
      \includegraphics[width=.45\textwidth]{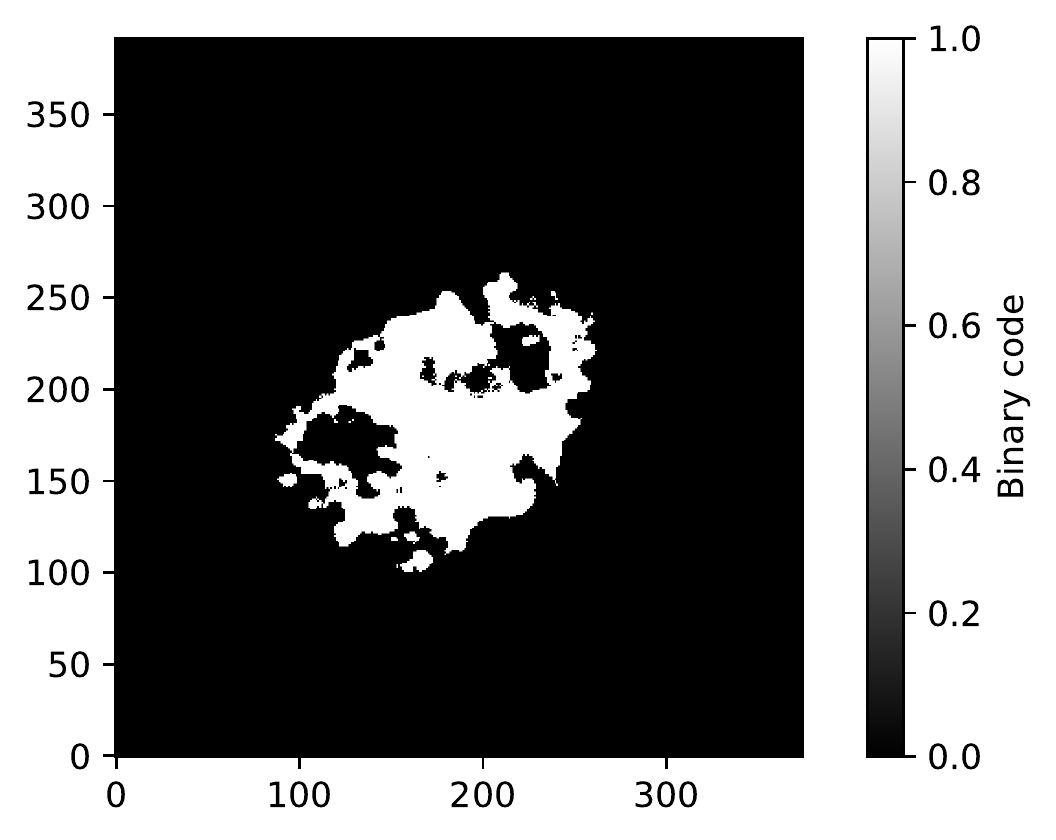}}
		
}
\vspace*{0.6cm}
\centering

   {\subfloat[After applying opening on the initial
mask]{  
      \includegraphics[width=.45\textwidth]{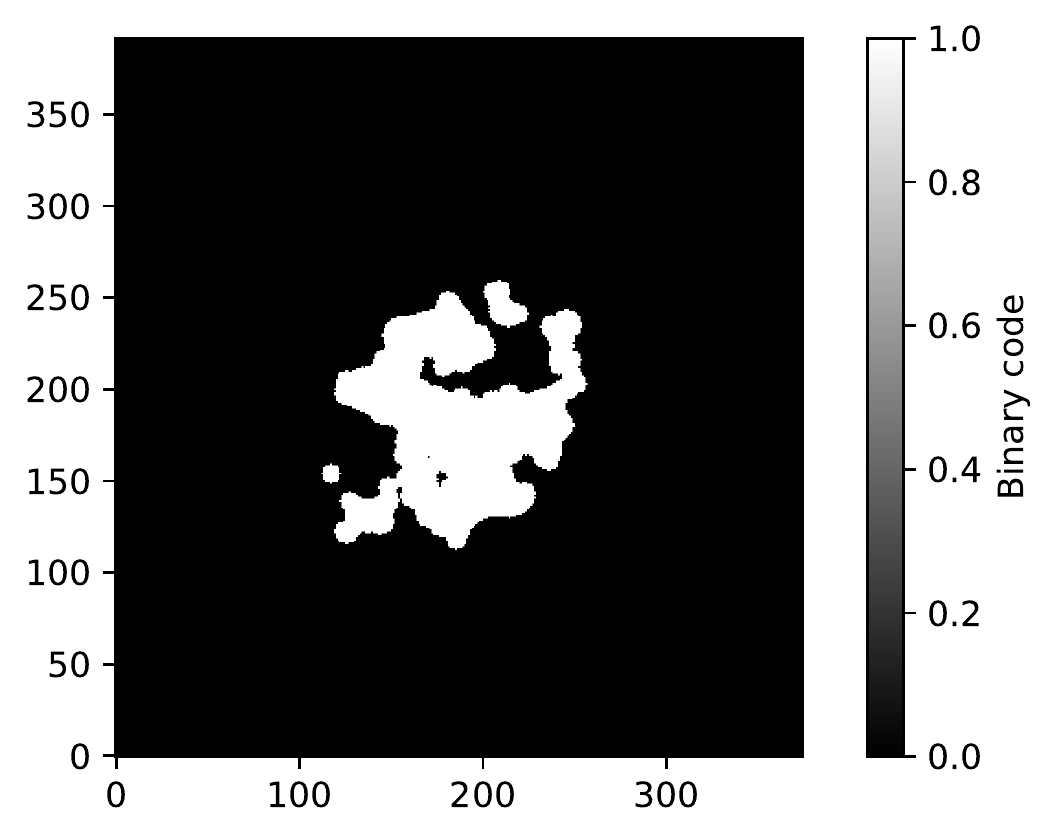}}
~
   \subfloat[Largest contour generated after
morphological processing
 ]{  \hspace*{0.6cm}
      \includegraphics[width=.45\textwidth]{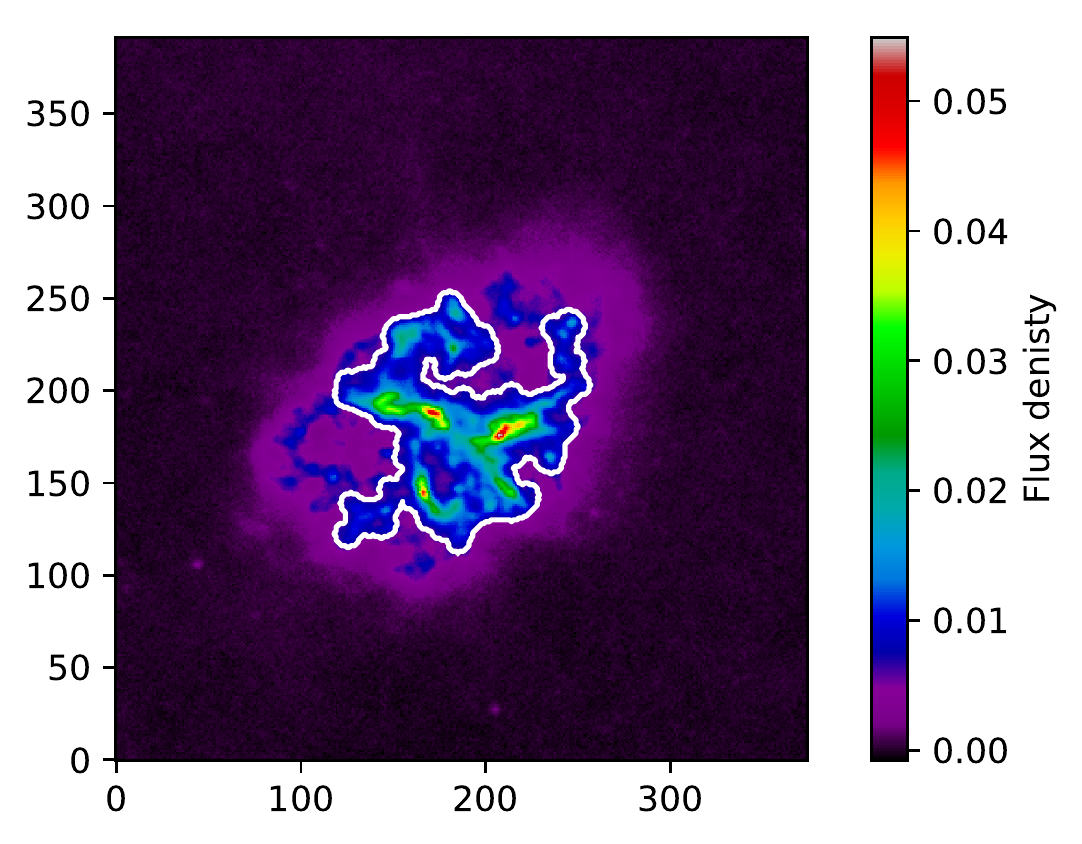}}
   }   
\caption{Initialization of the image processing procedure for $\lambda$=100\,\micron\ (PACS). Otsu's algorithm is applied on the frames of Fig~\ref{fig: zooms} to calculate to calculate an initial threshold value as represented in frame (a) and later a mask is generated based on that value as presented in frame (b). After applying \textit{opening} as shown in frame (c), isophotal contouring on the masked image is applied which will be later be used to find the pixel positions inside of it, frame (d).       
 \label{fig:otsu}}
\end{figure*}

To remove the effect of high spatial frequency noise which manifests itself in the form of detached or attached fine structures around the object of interest  in the masked image, we applied a morphological image processing step called \textit{opening} \citep[][see Fig.~\ref{fig:otsu},  b and c]{Morph}. After this step, we measured the length of all the contours that can be built on the binary masks, and we used the longest of them to define the region of interest (ROI) for photometry: all pixels with coordinates falling inside the contour (including the hollow regions excluded by Otsu's threshold, see Fig.~\ref{fig:otsu}c) contribute to the integrated flux associated to this value of the threshold. To each threshold value an ROI and thus a flux are associated, thus creating a functional relation from which the optimum threshold will be computed. The \textit{opening} step was necessary only for the PACS as the spatial noise is higher than in the SPIRE images. For wavelengths 70 and 100\,\micron, we used a circular morphological filters, also known as structural elements (strels), with a radius of 12 and 5 pixels respectively, and for 160\,\micron\ we used also circular strels but with a radius of 8 pixels. \textbf{The selection of the strels is essentially empirical. \textit{Opening} helped decreasing the number of contours that can be build on the binary mask at each step and, as a result, decreased the memory consumption and increased the speed of the process.}

\subsection{Finding the optimum threshold and integrated flux density}
\label{sec:find theta_op}

\begin{figure*}[htp]
\center
   {\subfloat[$2^{nd}$ iteration
]{  
      \includegraphics[width=.45\textwidth]{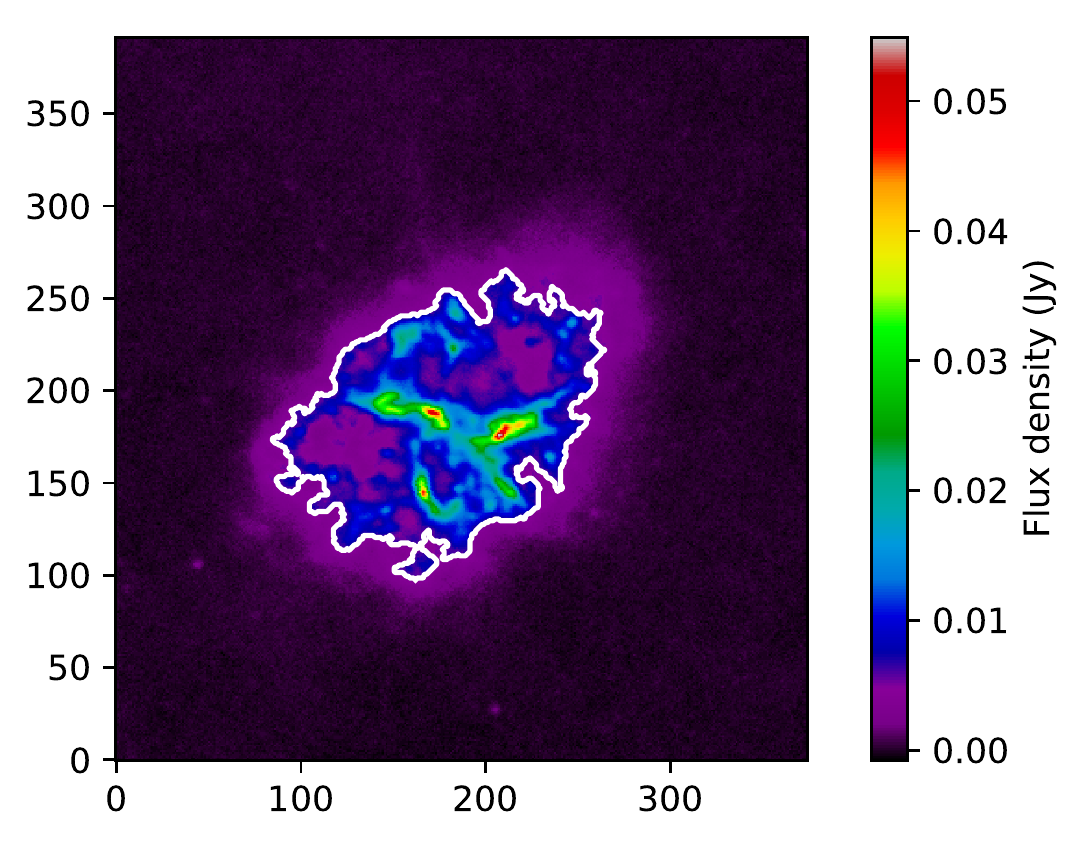}}
~
 \subfloat[$6^{th}$ iteration ]{  
      \includegraphics[width=.45\textwidth]{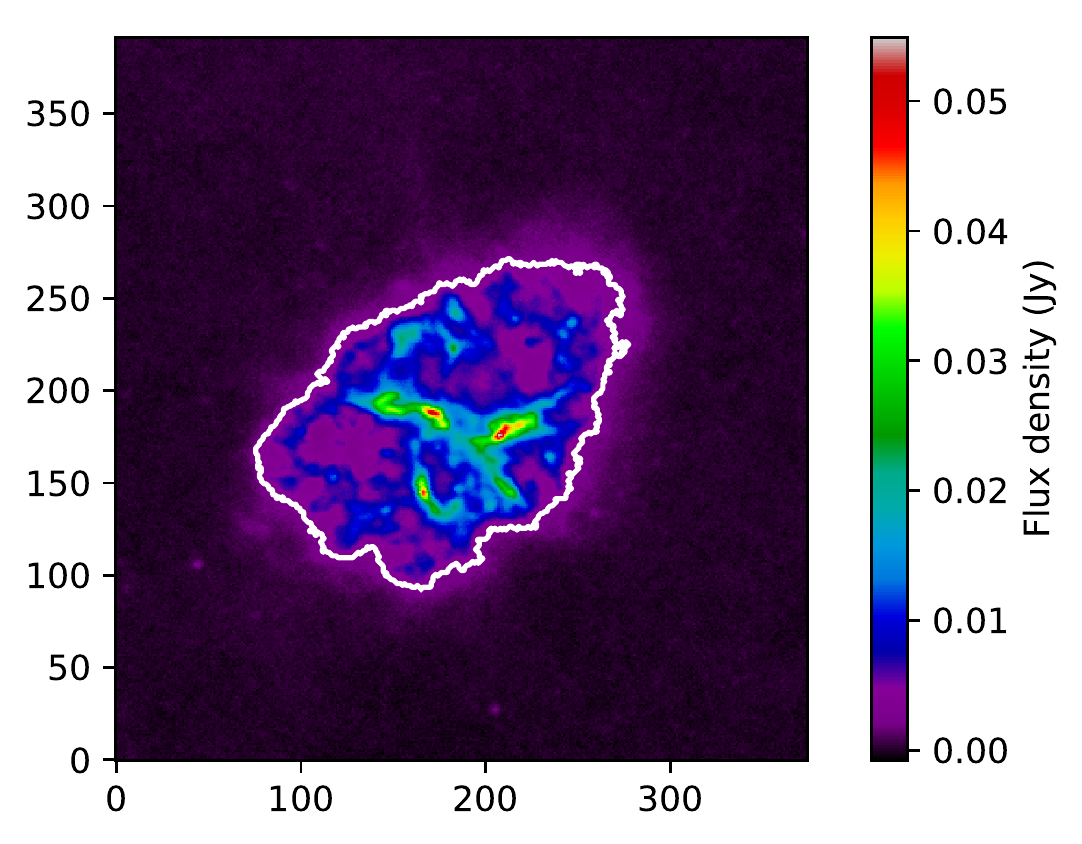}}

}

\center
   {\subfloat[$20^{th}$ iteration
 ]{  
      \includegraphics[width=.45\textwidth]{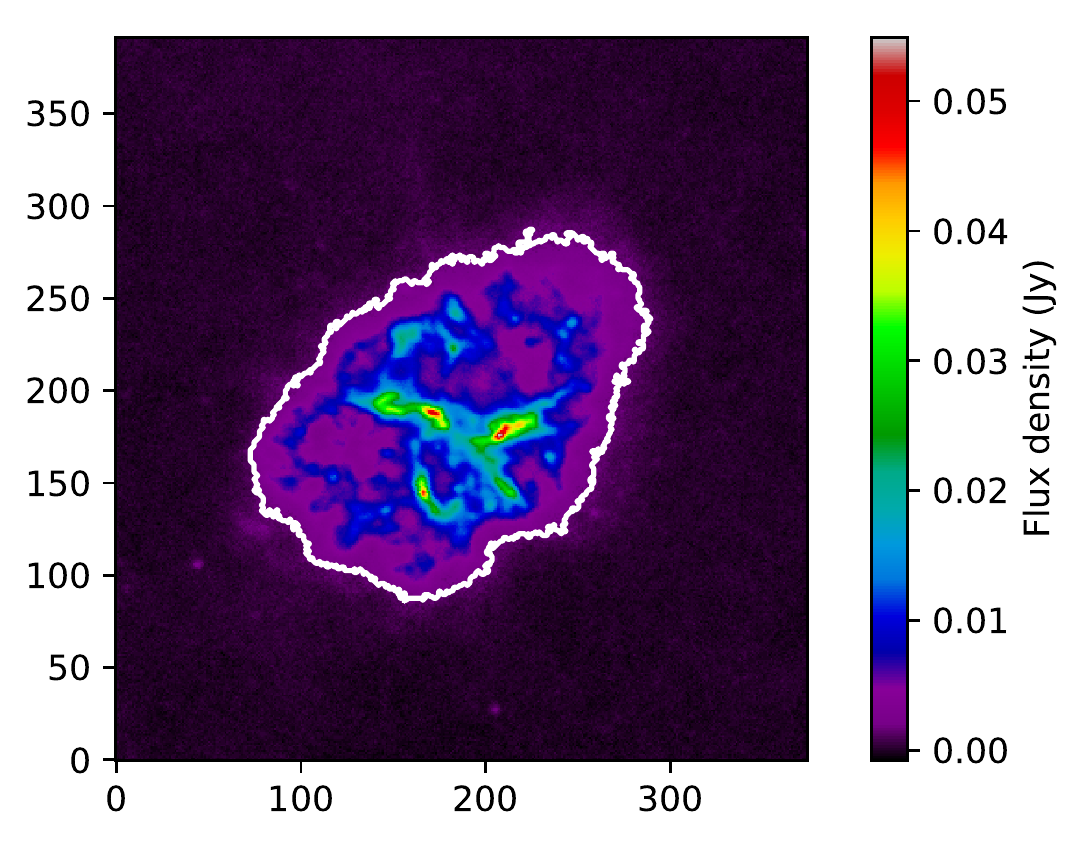}}
~
   \subfloat[$32^{nd}$ iteration ]{  
      \includegraphics[width=.45\textwidth]{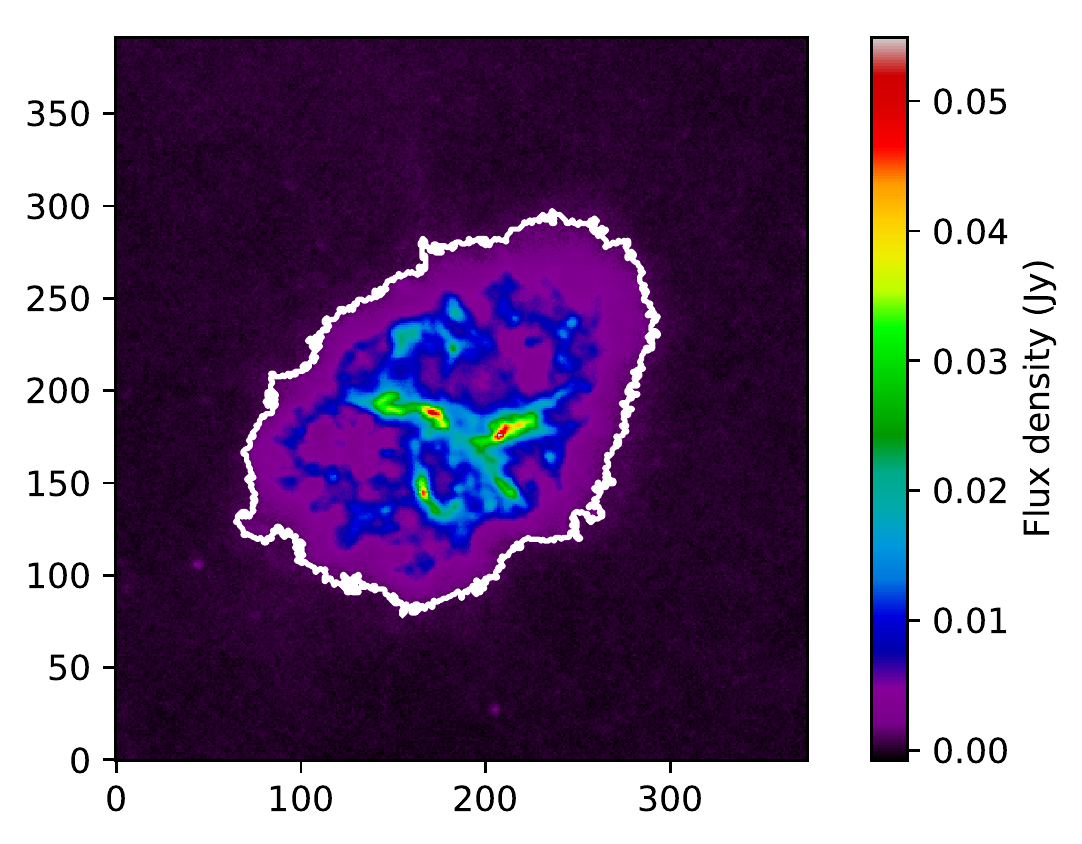}}
   }

   \caption{This is the image of $\lambda$=100\,\micron\ (PACS). In each iteration the threshold is decreased which causes more regions to be selected inside the isophotal contour.  
    \label{fig:iters}}
\end{figure*}

As the Otsu-derived threshold always clearly cuts into the object, iterations on the threshold consist in exploring decreasing values of it. During the descent of the threshold values, the generated ROIs will start to include a larger region of the Crab nebula and will include the outer regions where the flux density values tend to have a very slow and linear variation. The aim of this iterative approach is to objectively detect the edge of the nebula through characteristic features of the gradient of the integrated flux density. To this aim, we repeat the previously described process of ROI pixel coordinate generation for every threshold values, generating contours such as those displayed on Fig.~\ref{fig:iters}. As a result, we can build a piecewise functional relationship between  threshold values ($\theta$) and the integrated flux densities and can be defined as:

\begin{equation}
S_\nu^{tot}=f(\theta).
\end{equation}

The first degree numerical derivatives (gradient) is calculated because it is considered as a more reliable indicators to the optimum threshold: in some cases the $S_\nu^{tot}$ by itself would be sufficient to locate the optimum threshold, but in other cases the values were difficult to locate because of the nature of the variability near the edges of the object. For all the wavelengths however, the first derivatives did reveal sudden variations at the transition between foreground and background (see below). If $h$ is the step of $\theta$, then the numerical derivatives of $S_\nu^{tot}$ can be computed using the central differences in the interior:

\begin{equation}
S_\nu^{tot\, '}  \equiv\delta_h[S_\nu^{tot}](\theta)=S_\nu^{tot}(\theta+\frac{1}{2}h) - S_\nu^{tot}(\theta-\frac{1}{2}h),
\end{equation}

and the first differences at boundaries:

\begin{equation}
S_\nu^{tot\, '}  \equiv \Delta_h[S_\nu^{tot}](\theta)=S_\nu^{tot}(\theta+h) - S_\nu^{tot}(\theta),
\end{equation}

\begin{figure*}[htp]
\centering
\hspace{-0.5cm}
   {\subfloat[Integrated Flux density vs. threshold]{  
      \includegraphics[width=.5\textwidth]{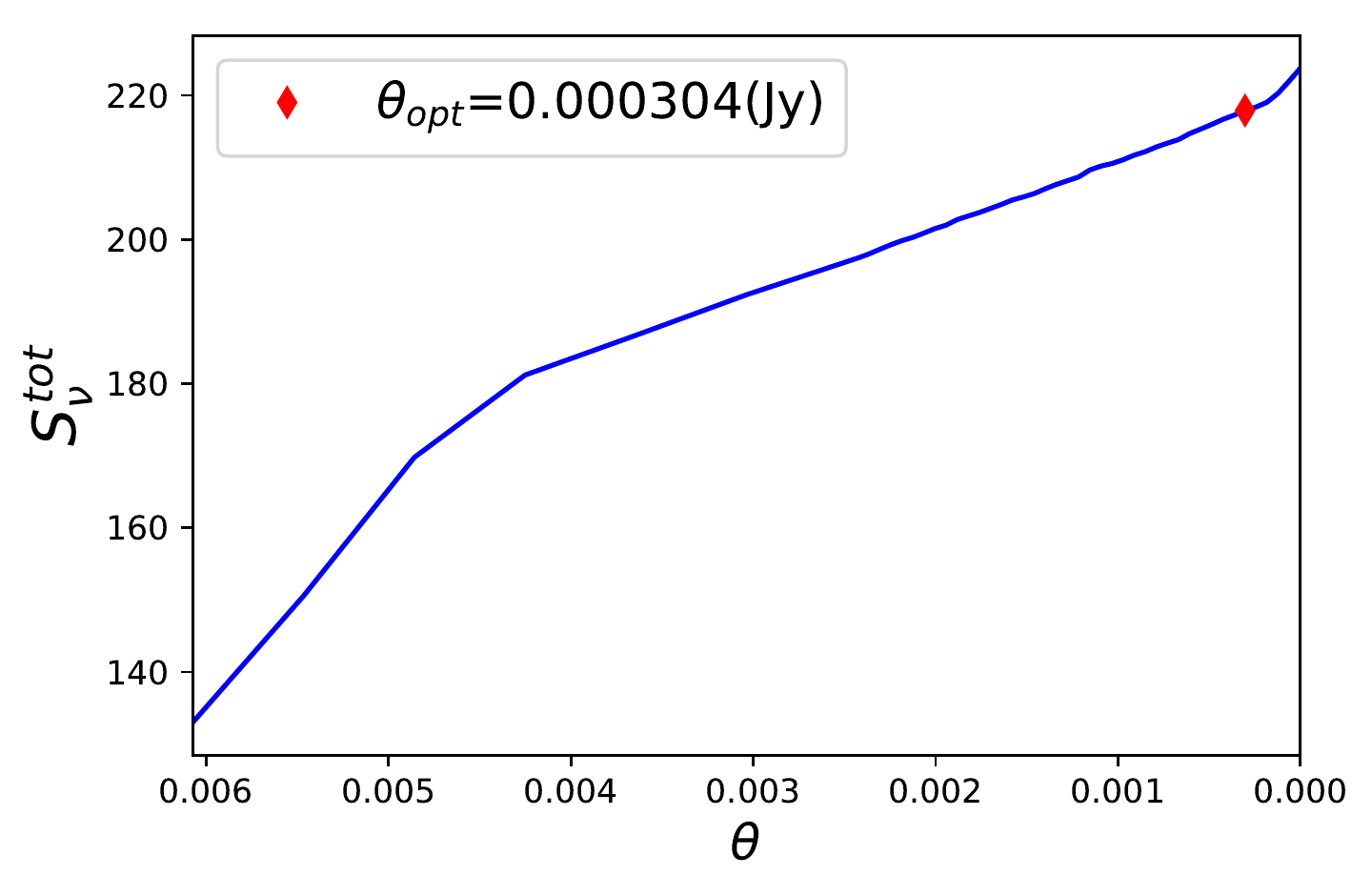}}
~
 \subfloat[Gradient of integrated flux density vs. thershold ]{  
      \includegraphics[width=.5\textwidth]{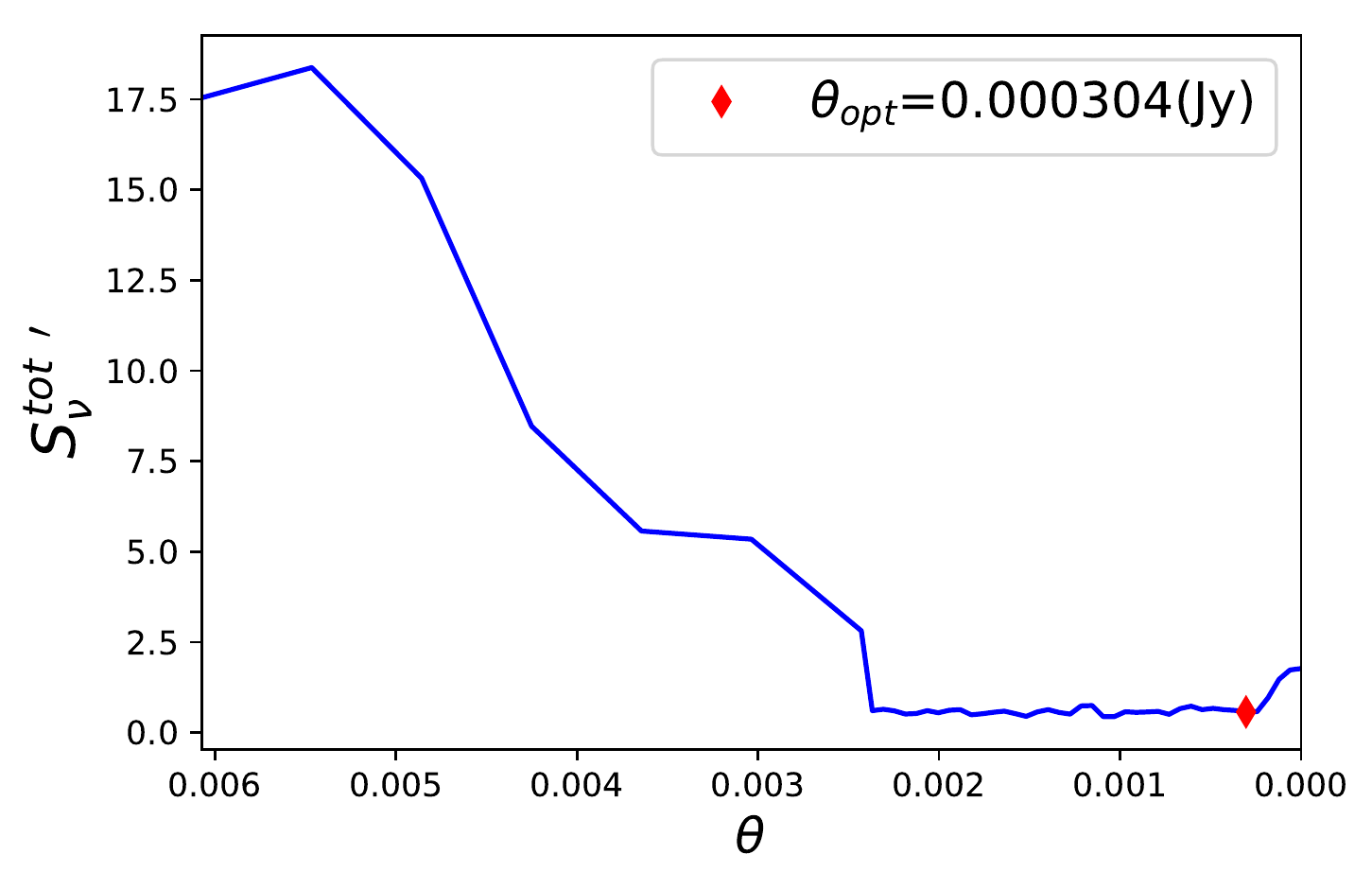}}

}

\centering
   {
   
\subfloat[Zoom in region of the crab nebula with the optimum ROI.]{  
      \includegraphics[width=.5\textwidth]{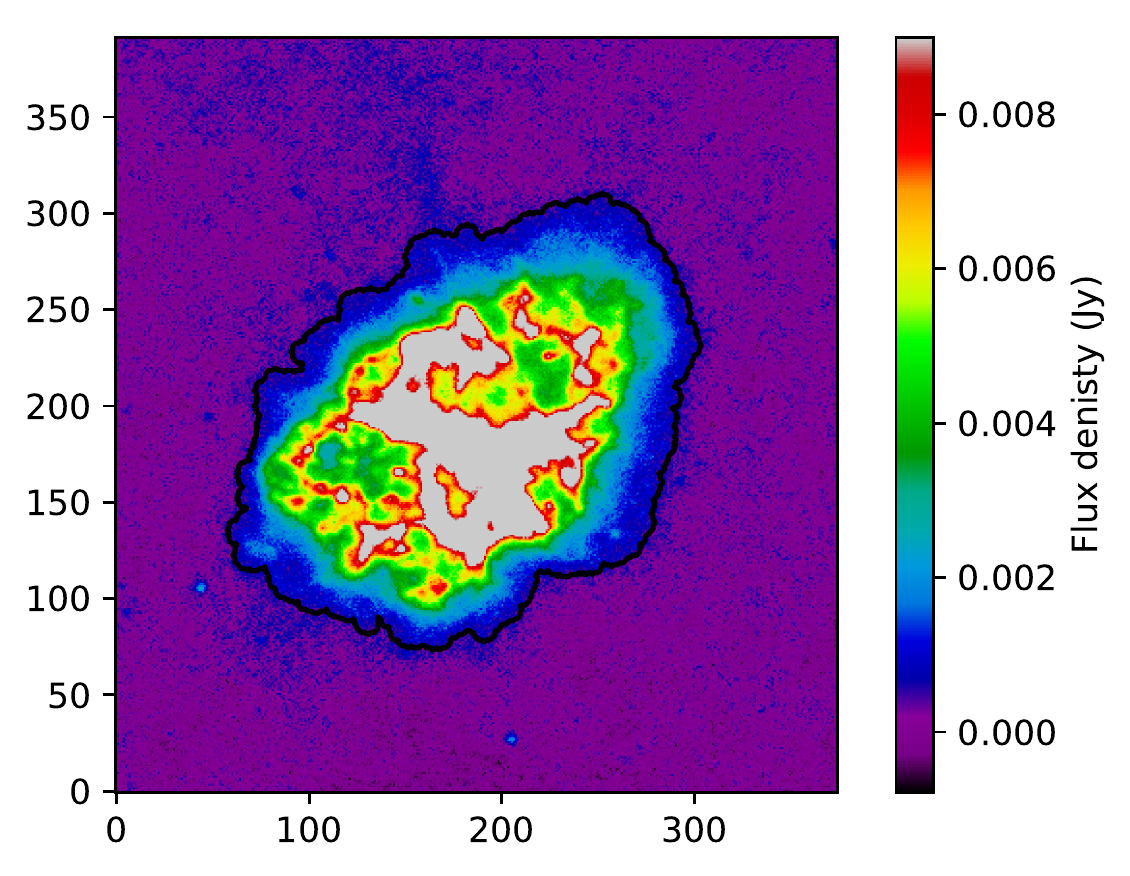}}

   }

   \caption{Images of $\lambda$=100\,\micron . The gradient of integrated flux density($S_\nu^{tot\,'}$ is used to detect the optimum ROI since the integrated flux density can't offer a direct indicator for it due to it linearly increasing nature. We present also in frame ``c" the isophotal contour of the optimum threshold. }\label{fig:plateau}
\end{figure*}

While decreasing the threshold values, the generated isophotal contours will start reaching the boundaries of the Crab nebula. Near the edges, these contours will start including pixel values which are lower (even negative) than that of the inner region of the Crab nebula. As a result, and in our tests, we realized that the $S_\nu^{tot}$ of ROI near the boundary with respect to $\theta$ will either have a linear shape or a sudden variation. When calculating the first order gradient, a constant value will be expected and we will see a plateau with a low signal to noise ratio followed by a sudden variation. The sudden variation following the plateau in the first derivative acts as the main indicator (see Fig.~\ref{fig:plateau}) for our optimum threshold.\\

To select the optimum threshold, the standard deviation of the neighborhood of each point of the first derivative was calculated. In our tests, a neighborhood of 6 step points and a standard deviation limit of 0.08 was used as a criteria for a threshold value to be considered belonging to the plateau. The smallest $\theta$ which passed this criteria was selected as the optimum threshold ($\theta_{opt}$). A posteriori visual inspection of the edges determined by this process revealed no issue that would lead to photometric errors, we therefore adopted it as our objective method to measure the Crab nebula's integrated flux density at all wavelengths (See Appendix for results and graphs).\\

\subsection{Derivation of the Nebula's integrated flux}

\begin{figure*}[htp]

\center
   {\subfloat[Modeling the background]{  
      \includegraphics[width=.45\textwidth]{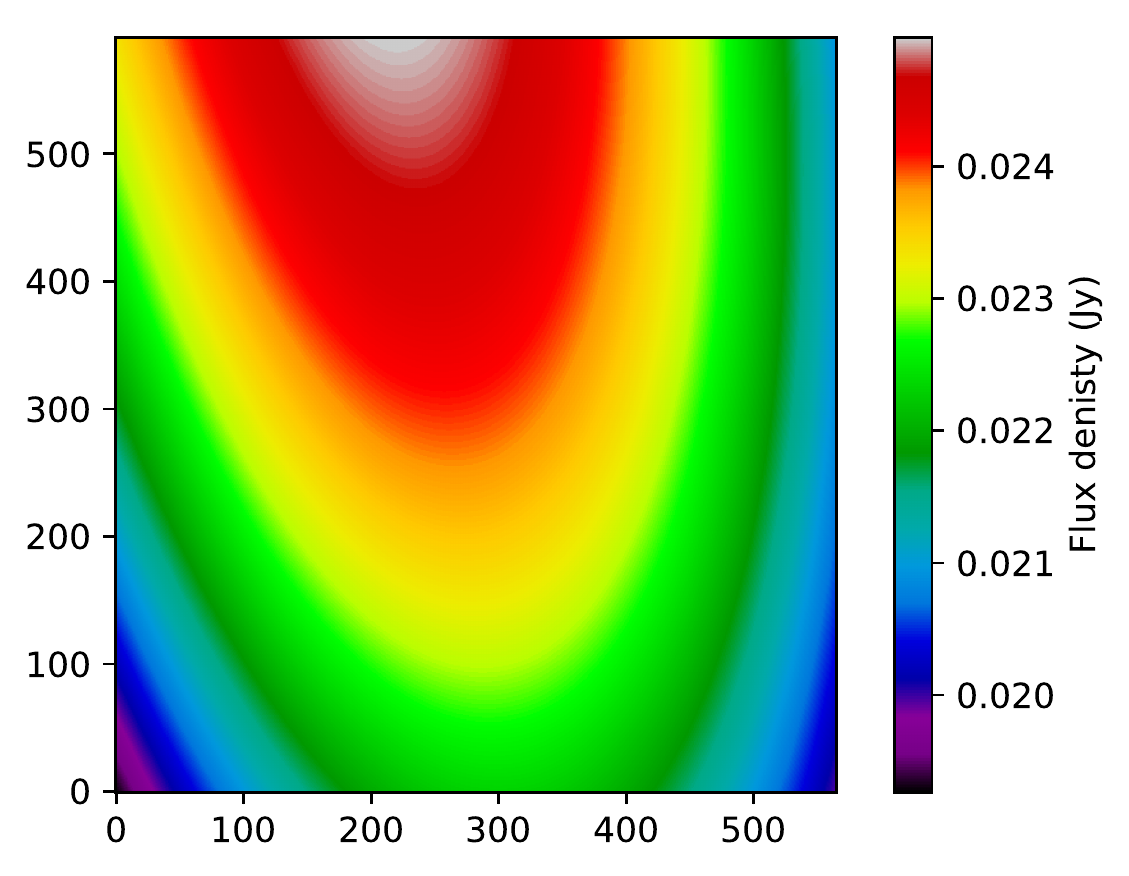}}
~
 \subfloat[The observation with the ROI masked out]{  
      \includegraphics[width=.45\textwidth]{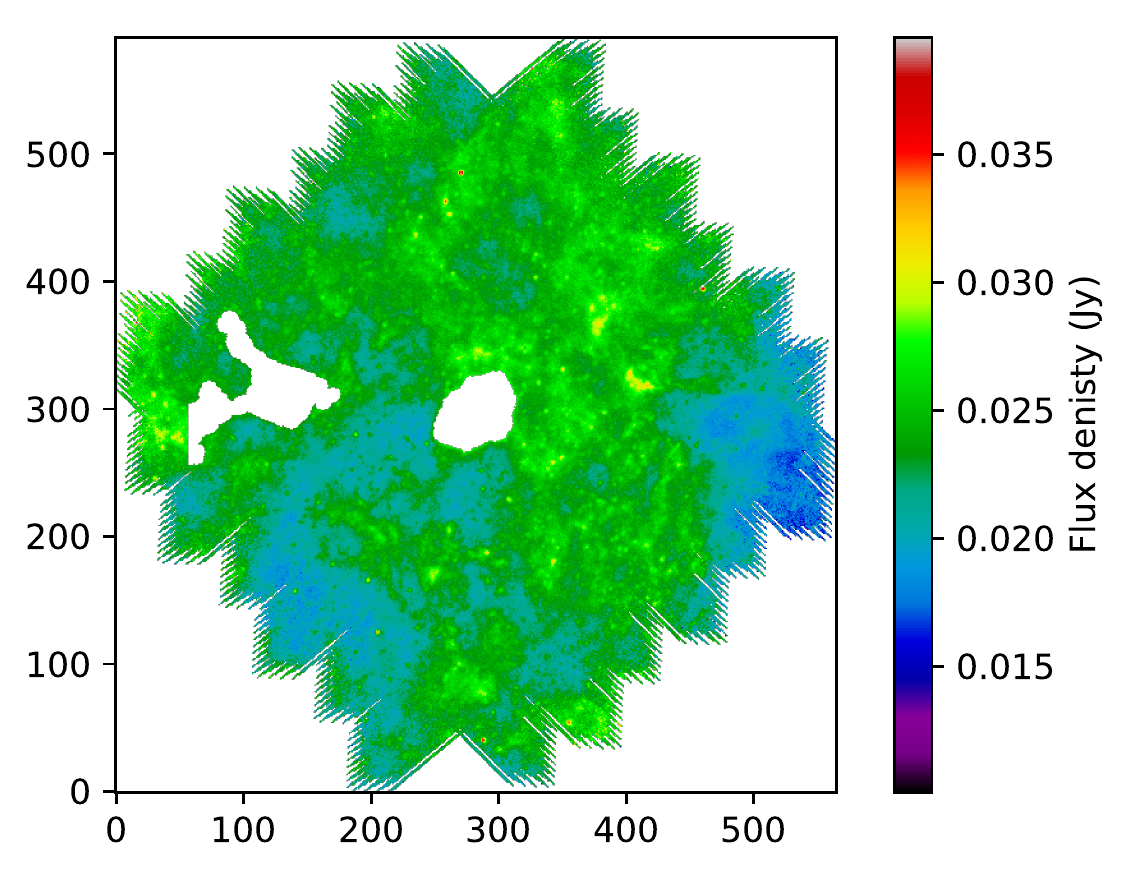}}

}

\caption{(a) an example of the resulting 2D polynomial model of the background, as derived from the fit to the masked $\lambda$=250\,\micron\ image presented in panel (b).}  \label{fig:model back}
\end{figure*}

With the object's ROI defined, the last step to determine flux of the object is to remove the background contribution in the ROI. We used the polygon generated by $\theta_{opt}$ at each wavelength and we used its interior pixel coordinates to mask the ROI. Then, a global surface interpolation is applied using a degree 2 polynomial on the pixels outside the ROI, {\bf to represent the large spatial frequencies in the background structure}, and the result of this interpolation is used to compute a model of the background at the same sampling and resolution as the original map (Fig.~\ref{fig:model back}). {\bf for clarity, this model image is referred to as $I_{back}$.}\\

A number of quantities attached to the ROIs can {\bf then} be derived. To establish a mathematical formalism, we assume that the $(x_i,y_i)$ are the coordinates of pixels, $I(x_i,y_i)$ is the {observed} flux density at each pixel, $\overline{I}$ is their average, $\alpha$ is the angular size of each pixel, N is the number of the pixels in the ROI. Hence we can define:\\  

The ROI area, to compare the sizes of the ROIs between different wavelengths:
\begin{equation}
A_{ROI}=N.\alpha^2. 
\end{equation}

{The total flux density in the optimum ROI:} 
\begin{equation}
S_\nu^{tot}= \sum_{i=1}^N I(x_i,y_i)_{ROI}.
\end{equation}

\textbf{The predicted background flux density in the optimum ROI:} 
\begin{equation}
S_\nu^{back}= \sum_{i=1}^N I(x_i,y_i)_{ROI}^{back}.
\end{equation}

The dispersion of the pixel fluxes in the optimum ROI:    
\begin{equation}
\sigma_{ROI}=\sqrt {1/N\sum_{i=1}^N \bigg(I(x_i,y_i)_{ROI}-\overline{I_{ROI}}\bigg)^2}
\end{equation}

The dispersion of the pixel fluxes outside the optimum ROI which traces the existence of small spatial structures superimposed on the large one fitted by the polynomial (the background area is composed of M pixels):    
\begin{equation}
\sigma_{back}=\sqrt {1/M\sum_{i=1}^M \bigg([I(x_i,y_i)-I(x_i,y_i)^{back}]-\overline{[I-I^{back}]}\bigg)^2}
\end{equation}


{\bf Further, since a polynomial is a crude representation of the complex background we are facing, we estimated a supplementary source of uncertainty: the area over which the polynomial fit is performed will impact the predicted level of the background in the center of the map, simply because a larger area potentially means a stronger deviation from our idealized polynomial representation of the background. We explored the variation of the predicted background contribution in the ROI ($S_\nu^{back}$) using areas that sample the map size beyond $\approx$3 times the object's size (we consider smaller sizes as being too close to the object to provide a fair representation of the complexity of inferring the background level ). From this we derived an estimated $\sigma_{str}$ which is an error component related to the large scale structures observed in the map. As can be seen in \ref{table:ROI flux}, the largest contribution is at 160\,\micron, where the edges of the map also suffer from data processing artefacts.

Finally, the combined error on the flux is derived as:

\begin{equation}
\Delta S_\nu=(\sigma_{ROI} + \sigma_{back} + \sigma_{str})\sqrt{N}
\end{equation}

We note that in principle, calibration errors enter the error budget. However calibration errors are systematic, and therefore apply simultaneously to all bands in an unknow similar way. They should therefore not be mixed with the statistical errors studied above. These error amount to 5\% in absolute photometry\footnote{see {\em Herschel} explanatory documentation on https://www.cosmos.esa.int/web/herschel/pacs-overview and https://www.cosmos.esa.int/web/herschel/spire-overview} and in effect contribute uncertainty at the mass determination level.

} 
 
The results of ROI calculations of the 6 images of PACS and SPIRE for wavelengths 70 to 500\,\micron\ are shown in Table~\ref{table:ROI flux}. Before moving on to the derivation of the infrared excess attributed to dust, we point out that our derived fluxes are compatible, within the respective error bars, with the previous measurements of \citet{gomez2012}, with the notable exception of the 500\,\micron\ flux. We also remark that our derivation is systematically lower. However we note that our flux derivation is (a) purely data driven, i.e. we let the data indicate in an objective manner how the object of interest should be extracted from the map, and (b) even though we proceed for each band independently we obtain an area for the object of interest which is quite constant (and slightly larger at 500\,\micron, thus indicating that our lower flux value is not due to a smaller aperture).
This will obviously have an impact when computing the dust mass in the nebula. 
{\bf To further increase our confidence in our photometric process, we have explored how the background-corrected object flux vary if we artificially increase the ROI size by an amount similar to the observed dispersion of sizes as a function of wavelength (column 4 in Table~\ref{table:ROI flux}), or greater as, for the longer wavelengths, expanding the perimeter by just one pixel already leads to a larger area increase than the dispersion. The observed flux variations are significantly smaller that the flux uncertainties listed in the table. Therefore the observed area variations from one band to another have no impact on the derived fluxes, and we postulate that differences with previous measurements indicate that the latter have been contaminated by foreground/background emission at the longest wavelengths.}\\

\subsection{Constructing the SED}

After calculating the integrated flux densities, we continue constructing the SED of the Crab nebula. Due to the pulsar, the SED is dominated by the synchrotron radiation, on which is superposed an IR flux excess corresponding to the dust emission that we wish to model, to quantify the mass of dust in the remnant. Therefore it is necessary analyze the synchrotron component and remove its contribution. The synchrotron flux density model can be defined as:
\begin{equation}
\label{eq 5}
S^{syn}_{\nu} = A\lambda^{\gamma}.
\end{equation}

\begin{figure}
\centering
\includegraphics[scale=0.55]{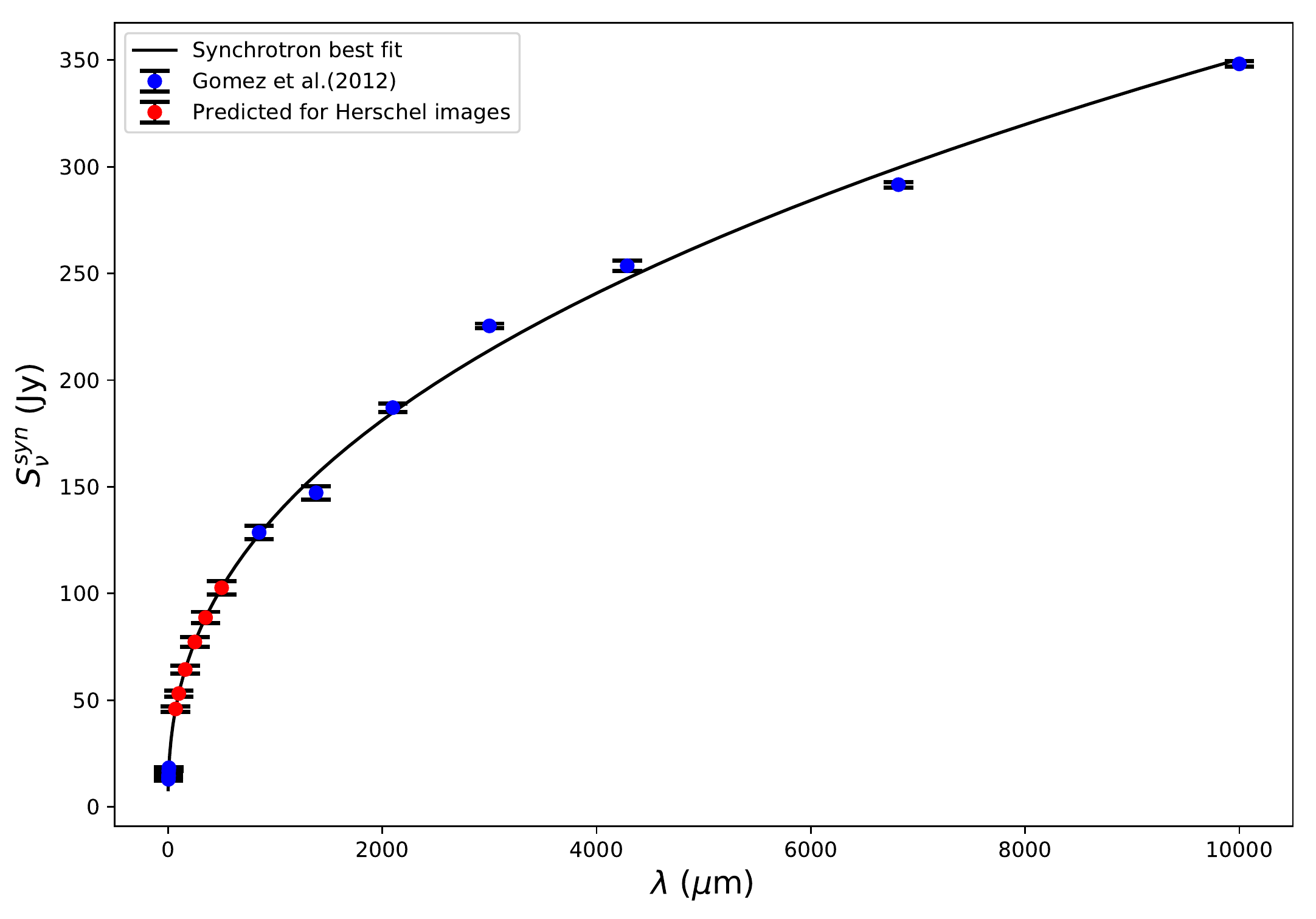}
\caption{Modeling the synchrotron emission using a power law using the results of \cite{gomez2012} and predicting the synchrotron flux densities for SPIRE and PACS images.}
\label{fig:synchrotron fitting}
\end{figure}

Thus to achieve this step, we applied a power law fit by minimizing a merit function ($\chi^2$) with respect to the spectral index ($\gamma$) and the amplitude of flux density ($A$). To account for the measurement errors, the merit function $\chi^2$ was weighted to handle the uncertainties. The uncertainty of the flux density at each wavelength was considered as the error weight ($\sigma_i$). Thus the weighted merit function becomes :
\begin{equation}
\label{eq 8}
\chi^2_w=\sum_{i=1}^N \Bigg( \frac{S^{syn}_{\lambda_i}-y_i}{\sigma_i}\Bigg)^2,
\end{equation}
where $S^{syn}_{\lambda_i}$ is the synchrotron flux density model evaluated at the $i^{th}$ band, $y_i$ is the observed data point. The data used for modeling the synchrotron radiation was adopted from \cite{gomez2012}, excluding the {\em Herschel} bands to avoid contaminating the fit by dust emission (see Table~\ref{table: synchrotron data} and Fig.~\ref{fig:synchrotron fitting}).\textbf{ We have also acknowledged the existence of the spectral break value derived by \cite{gomez2012} at 10\,000\,\micron, and applied the fitting up to that point.  }

 The power law fit resulted in a spectral index $\gamma=0.410\pm0.003$ and an amplitude equal to $A=8.039 \pm 0.195$\,Jy.\micron$^{-\gamma}$.

\begin{table}[htp!]
\centering

\begin{tabular}{p{.03\textwidth}  p{.03\textwidth}  p{.04\textwidth}  p{.065\textwidth}  p{.065\textwidth}  p{.065\textwidth}  p{0.065\textwidth} p{.065\textwidth} p{.05\textwidth} p{.05\textwidth} p{.05\textwidth} p{.09\textwidth}}

\hline
\hline\\
$\lambda$ &   $\alpha$&  $N$ &   $A_{ROI}$ &   $S_\nu^{tot}$  & $S_\nu^{back}$ &   $S_\nu$ &   $\Delta S_\nu^{ROI}$ &   $\Delta S_{\nu}^{back}$ &  $\Delta S_{\nu}^{str}$ &   $\Delta S_{\nu}$&   $\theta_{opt}$  \\

(\micron) &   (")& (pix)  &   ("$^{2}$) &  (Jy) &  (Jy) &   (Jy) &   (Jy) &   (Jy) &   (Jy) &   (Jy) &   (Jy)  \\

(1) &   (2)& (3) &   (4) &  (5) &  (6) &   (7) &   (8) &   (9) &   (10) &   (11) &   (12)  \\

\hline
\hline\\

  70 &  1.6 & 53858 & 137876  & 219.509 &  4.491 & 215.017 & 1.380 &  0.055 &  0.336 &  1.771 & -0.000302 \\
  100 &  1.6 & 39920 & 102195 & 217.895 &  8.805 & 209.090 & 1.165 &  0.049 &  1.467 &  2.681 &  0.000303 \\
  160 &  3.2 &  9660 &  98918 & 140.946 &  9.940 & 131.007 & 1.111 &  0.090 &  8.902 & 10.104 &  0.00275 \\
  250 &  6 &  2910 & 104760 & 162.376 & 69.860 &  92.516 & 1.246 &  0.106 &  1.257 &  3.347 &  0.0277 \\
  350 & 10 &  1186 & 118600 & 144.968 & 48.730 &  96.238 & 2.224 &  0.104 &  1.019 &  5.059 &  0.0476 \\

  500 & 14 &   697 & 136612 & 127.976 & 24.716 & 103.260 & 3.448 &  0.069 &  1.538 &  8.433 &  0.0450 \\

\hline
\hline
\end{tabular}
\caption{Results of calculation for all the images. The first column lists the reference wavelength of each photometric band. It is clear that as the resolution of the image increases (and pixel size decreases to maintain adequate sampling), the number of number of pixels inside the ROI is increasing as well, however the area itself is quite constant,\textbf{ statistically it amounts to  $116493.5\pm2\,650.5''^2$, where the uncertainty is computed as {\footnotesize 1/6$\times\sqrt{1/6\sum_{i=1}^6 (A_{ROIi}-\overline{A_{ROI}})^2}$.}}}
\label{table:ROI flux}
\end{table}

After calculating the power law fit, we predict the synchrotron emission flux densities and their uncertainties for PACS and SPIRE wavelengths (see Table~\ref{table:IR excess}, columns 4 and 5).  To calculate the excess IR, we subtract the background and synchrotron contributions from the observed total flux:
\begin{equation}
S_{\nu}^{IR}=S_{\nu}^{tot}-S_\nu^{back} - S^{syn}_{\nu}\,.
\end{equation}

\begin{table}[!h]
\centering
\begin{tabular}{ccc}
\hline
\hline\\ \vspace*{0.2cm}
   $\lambda$(\micron) &   $S^{syn}_{\nu}$(Jy) &   $\Delta S^{syn}_{\nu}$(Jy) \\
   
\hline   
\hline\\
     3.4 &     12.9 &      0.6  \\
     4.5 &     14.4 &      0.26 \\
     4.6 &     14.7 &      0.75 \\
     5.8 &     16.8 &      0.1  \\
     8   &     18.3 &      0.13 \\
   850   &    128.6 &      3.1  \\
  1382   &    147.2 &      3.1  \\
  2098   &    187.1 &      2    \\
  3000   &    225.4 &      1.1  \\
  4286   &    253.6 &      2.5  \\
  6818   &    291.6 &      1.3  \\
 10000   &    348.2 &      1.2  \\

\vspace*{0.2cm}      
       10000   &  348.2 &    1.2  \\

\hline
\hline
\end{tabular}
\caption{Flux densities used to model the synchrotron SED using data from \cite{gomez2012}. We have specifically avoided using bands located too close to the infrared excess.}
\label{table: synchrotron data}
\end{table}

The results of the flux density calculations are presented in Table~\ref{table:IR excess} (columns 6 and 7).  As hinted before, we measure no excess due to dust at 500\,\micron.\\

\begin{table}[!t]
\centering
\begin{tabular}{ccccccc}
\hline
\hline\\

   $\lambda$(\micron) &  $S_\nu$(Jy) &   $\Delta S_\nu$(Jy) &  $S^{syn}_{\nu}$(Jy) &   $\Delta S^{syn}_{\nu}$(Jy)& $S_\nu^{IR}$ (Jy)&$\Delta S_\nu^{IR}$(Jy) \\
   
\vspace*{0.2cm}
(1)& (2)& (3)& (4)& (5)& (6)& (7) \\    
   
   \hline
   \hline\\ 
   
70 & 215.02 & 1.77 &   45.86 &    1.26 & 169.16 &    3.34 \\
100 & 209.09 &  2.68 &   53.08 &    1.49 & 156.01 &    4.08 \\
160 & 131.01 & 10.10 &   64.36 &    1.86 &  66.65 &   11.28 \\
250 &  92.52 &  2.61 &   77.27 &    2.29 &  15.24 &    4.39 \\
350 &  96.24 &  3.35 &   88.70 &    2.68 &   7.54 &    6.09 \\
\vspace*{0.2cm}
500 & 103.26 &  5.06 &  102.66 &    3.17 &   0.60 &    9.01 \\
      
\hline
\hline
\end{tabular}
\caption{This table summarizes that flux density calculation and the relevant error bars.}
\label{table:IR excess}
\end{table}

\section{Dust grain model}
\label{sec:dust}

The Crab Pulsar (B0531+21) is the brightest source of synchrotron radiation in our
Galaxy. Its emission  in the NIR and MIR is dominated by continuum synchrotron radiation. The dust emission appears as an excess superimposed over that synchrotron continuum, with dust grains emitting, to a first approximation, a modified back-body spectrum (hereafter MBB, the black-body spectrum modified by the spectral dependence of the dust absorption cross section). For that emission, the basic parameters necessary to describe the emission from dust grains are the temperature T$_d$ and a form of the emissivity function $\epsilon_{\nu}$.\\

 At FIR wavelengths ($\lambda \geq 60$\,\micron), the shape of the SED due to thermal emission from dust is empirically found to be well represented by the Planck function:

\begin{equation}
\label{eq BB}
B(\nu,T)=\frac{2h\nu^3}{c^2}\frac{1}{e^{\frac{h\nu}{k_BT}} - 1},
\end{equation}

which is evaluated at the dust temperature T, modified by a power law frequency dependent $\nu^{\beta}$ \citep{Casey2012}. More precisely the dust emission flux density can be written as:

\begin{equation}
S_{\nu}^{IR} = \Omega N \kappa_0 \left( \frac{\nu}{\nu_0} \right)^{\beta} B(\nu,T),
\label{equation:mbbfull}
\end{equation}

\noindent where $\Omega$ is the solid angle of the observing beam, N is the column density of emitting material, $B(\nu,T)$ is the Planck function, and $\kappa_0 \left( \frac{\nu}{\nu_0} \right)^{\beta}$ is the opacity of the emitting dust. Note that $\tau_0 = N\kappa_0$ is the optical depth at frequency $\nu_0$. An essential assumption in Equation (\ref{equation:mbbfull}) is that dust is optically thin, so that $\tau(\nu) \ll 1$ \citep{Kelly2012}, and hence all the dust grains within the beam contribute emission. Dropping the constants from this equation gives:

\begin{equation}
S_{\nu}^{IR} \propto \nu^{\beta} B(\nu,T).
\label{equation:mbbsim}
\end{equation}

The spectral index $\beta$ determines the opacity $\kappa_{\nu}$ of the dust, and encodes information about grain composition. It is found to be in the range 1 to 2 for the silicate and/or carbonaceous grain composition common in the diffuse ISM \citep{Draine1984}. Drawing from \citet{hildebrand1983}, the dust mass absorption coefficient is:

\begin{equation}
\label{eq kappa}
	\kappa_{\nu} = K_{3000GHz}(\frac{\nu}{3000\,\rm{GHz}})^{\beta},
\end{equation}
\textbf{where for $\beta$ = 1.5, $\kappa_{3000GHz}=40$ cm$^2$ g$^{-1}$.\\
}

The IR excess attributed to dust in the Crab nebula {\bf can adequately be} modeled by {\bf a single MBB}. From this component we deduce from the SED fit a temperature and the dust mass $M_{d}$. Starting from equation (\ref{equation:mbbfull}), and relating the solid angle and column density to the distance of the object and total mass of grains \citep[see for instance][]{Laor1993,Weingartner2001} we get:
\begin{equation}
S_{\nu}^{IR} = \frac{M_d \kappa_{\nu}}{D^2}  B(\nu,T),
\label{Md}
\end{equation}
where, $S_{\nu}^{IR}$ is the flux density, $\kappa_{\nu}$ is the dust mass absorption coefficient, and $D$ is the distance between to the Crab Nebula and is equal to $2.0 \pm 0.5$ Kpc \citep{Trimble1973}. Combining equations
(\ref{eq BB}), (\ref{eq kappa}), and (\ref{Md}) we will get:

\begin{equation}
\label{eq model}
S_{\nu}^{IR} = \frac{8\times10^{26}}{D^2} \frac{h\nu^3}{c^2}\bigg(\frac{\nu}{3\times10^{12} Hz}\bigg)^{1.5} \, \Bigg[\frac{M_{d}}{e^{\frac{h\nu}{k_BT}} - 1} \Bigg].
\end{equation}

Where now the equation is rearranged to be in Jy. In the modeling the optimization process will be based on the main two parameters which are the mass and the temperature. Statistically speaking, while optimizing the $\chi^2$, the predictor will be the frequency ($\nu$ in Hz) and the response will be the integrated flux density ($S_{\nu}^{IR}$ in Jy)\\

\section{IR excess SED fitting and dust mass calculation }
\label{sec:irex}

\begin{figure}[htp!]
\centering
\includegraphics[width=15cm]{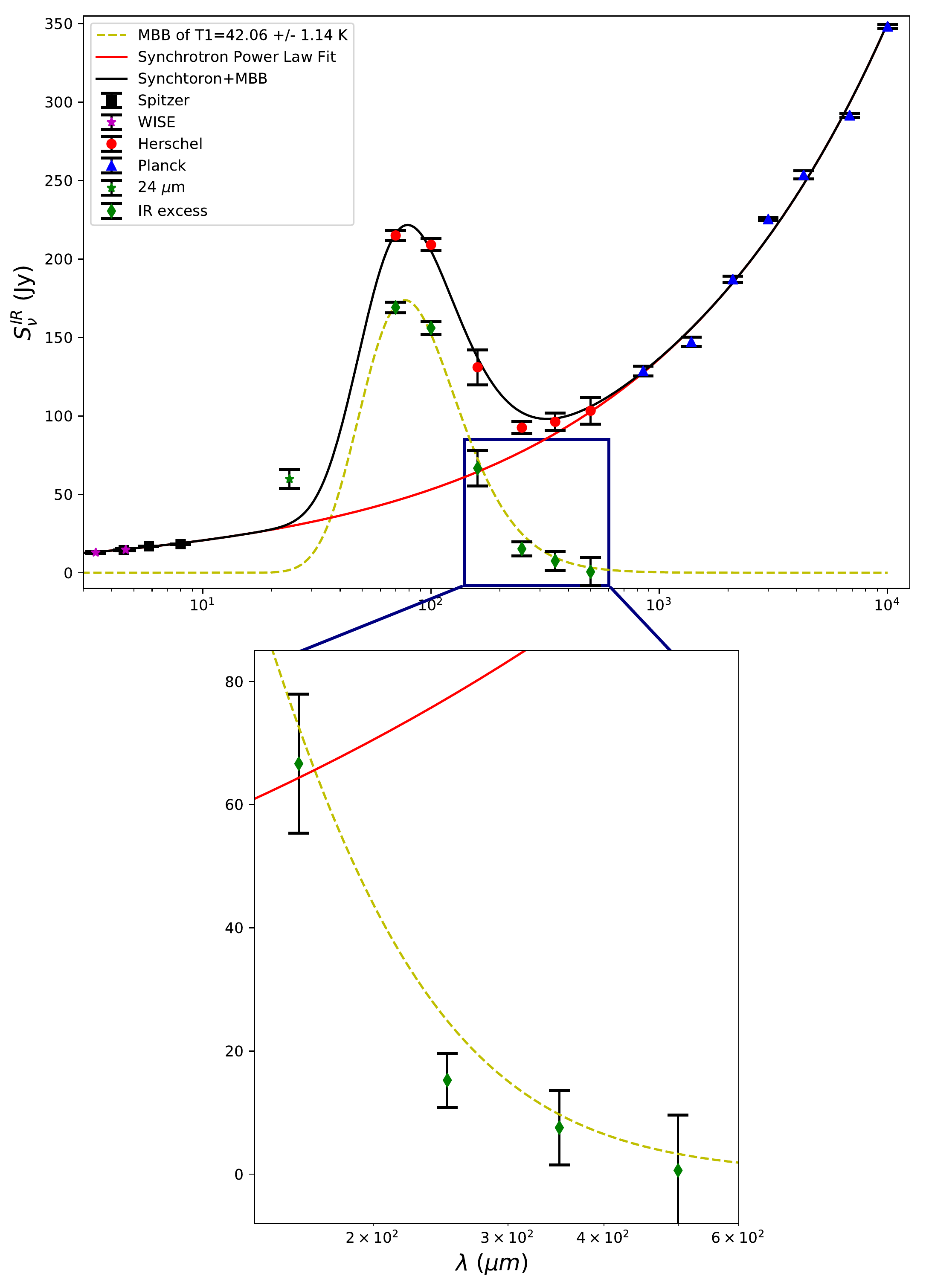}
\caption{Semi-log scale plot of the SED. The zoom on the $250\,\mu$m region shows that it has a higher deviation compared to the other data points.}
\label{fig:semilog_zoomed} 
\end{figure}

\begin{figure*}[htp!]
\centering

\hspace*{-1.5cm}
\includegraphics[width=17cm]{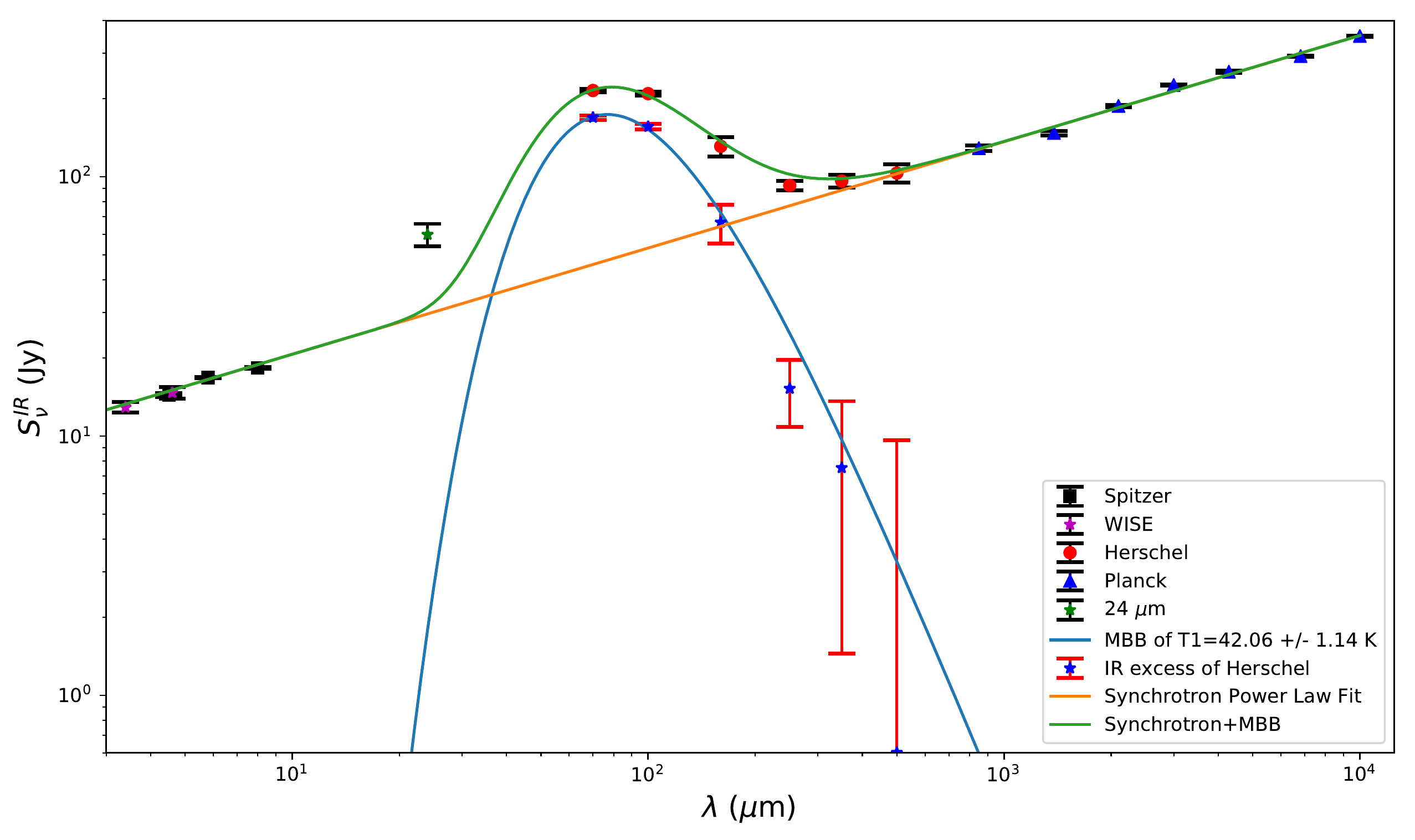}
\caption{log-log scale plot of the SED showing the fitting process.  }
\label{fig:loglog} 
\end{figure*}

Although the modeling is based on $\nu$, in Figs.~\ref{fig:loglog} and ~\ref{fig:semilog_zoomed} the plots of the SED of the Crab nebula and the results from the fitting process are presented in a more traditional way as a function of the wavelength ($\lambda$). In both figures, we can see the fitting of the IR excess, in addition to a best fit to which the synchrotron flux is added. The results for the dust mass computation are presented in Table ~\ref{table:mass calc}. We used the model-fitting open-source \texttt{scipy} package \texttt{curve\_fit()} which applied fitting by including the uncertainties of the values of $S_{\nu}$\footnote{The description of the function with the related computation and references are explained in the documentation of the package: https://docs.scipy.org/doc/scipy/reference/generated/scipy.optimize.least{\textunderscore}squares.html\,.}.\\

\begin{table}[!h]
\centering
\begin{tabular}{ccc}
\hline
\hline\\
\vspace*{0.2cm}

  Parameter &  value $\pm$ $\Delta$ value \\  
  
\hline
\hline\\  
    
$T$ & 42.06 $\pm$ 1.14 (K)  \\

$M_{d, \Delta D=+0.5Kpc}$ & 0.084 $\pm$ 0.010 (\Mo)   \\

$M_{d, \Delta D=-0.5Kpc}$ & 0.030 $\pm$ 0.004 (\Mo) \\

$M_{d, \Delta D=0Kpc}$ & 0.054 $\pm$ 0.007 (\Mo)\\

$\overline{M_{d}} $ & 0.056 $\pm$ 0.034 (\Mo)\\
\vspace*{0.2cm}
$\overline{M_{d,cal=5\%}} $ & 0.056 $\pm$ 0.037 (\Mo)\\

\hline
\hline
\end{tabular}
\caption{The characteristics of the dust based on the single component modified black-body model.}
\label{table:mass calc}
\end{table}

In our modeling, we assumed the distance to the Crab nebula to be known, however to provide estimates of the mass, we calculated it using an upper and lower boundary values for the distance, since for a given SED, the mass and the square of the distance are proportional to each other. We calculated mass average of the mass  $\overline{M_{d}}=0.056$ \Mo\, and the uncertainty as the distance from the mean to the boundary values.
The total uncertainty on the mass is sum of fitting and calibration errors (5\%), or 0.037\,\Mo.\\

{\bf As can be seen on Figure~\ref{fig:semilog_zoomed}, the fit is poorer at 250\,\micron. We see no indication in our photometric process that this particular flux derivation would be incorrect. We have investigated whether a different assumption on the dust properties, in particular the exponent $\beta$ of the emissivity could provide a better fit to the SED, and indeed a better fit is obtained with $\beta=2$. While this can be instructive, we still keep to our initial selection of a value of 1.5 as it allows direct comparison with previous work, and is in effect tied to the absolute value of the mass absorption coefficient that we have used.}

\section{Discussion and conclusion}
\label{sec:concl}

We have presented an improved method for calculating the flux densities using several image processing techniques applied on PACS and SPIRE images.  With an optimal object-threshold selection method we constructed the spectral energy distribution of the Crab nebula in the {\em Herschel} band. We modeled the synchrotron component and removed its contribution to extract the IR excess. {\bf The resulting SED can adequately be modeled} with a single component MBB using weighted least square fitting. {\bf This leads to }a dust component of parameters $T=42.06\pm1.14$\,K and $\overline{M_{d,cal=5\%}}=0.056\pm0.037$\,\Mo. The combined (synchrotron+dust) SED is shown in Fig~\ref{fig:loglog} and ~\ref{fig:semilog_zoomed} where we see that it under-predicts the flux in the range around 20\,\micron, i.e. it does not violate the constraint set by the Spitzer observation at 24\,\micron\ \citep[$S_\nu=59.8\pm6.0$ Jy; adopted from][]{gomez2012}. As mentioned earlier, we consider that using a thermal process in this wavelength range is not justified, and remark that thermally fluctuating dust particles (responsible for emission in that range) will anyway not affect much the total mass in the nebula.  \\

{\bf Our objective in this paper is primarily to estimate how the photometric methods impact the dust mass determination. We thus now} compare our results with previous determinations. \citet{gomez2012} and \citet{tea2013} have provided reference derivation of the dust content of the nebula, where the main differences lies in the dust model that was used to convert from emission to mass. Our work is most comparable to that of \citet{gomez2012} as \citet{tea2013} used a more complex approach involving a simplified radiation transfer approach that led to temperature distributions for the dust component. While our value of the dust mass is comparable to the carbon-based single component model of \citet{gomez2012}, where the temperature is 40\,K and the mass is 0.08\,\Mo, there is a noticeable difference in the fact that the single component model is not the favored result of \citet{gomez2012}. Rather they favor a two-component model with temperatures bracketing our best-fitting temperature, and as a result a total dust mass 2 to 4 times larger due to the presence of cold dust. Two facts explain this difference: (1) we chose not to include observations below 70\,\micron\ in the fit for the reasons explained above, and (2) our dust excess SED is significantly different from that of \citet{gomez2012}, and in particular, with no excess at 500\,\micron, leaves no room for a cold component. {\bf This stresses how complex and uncertain the process of determining the dust content in the Nebula is: cold dust easily becomes the dominant component in mass but rests mostly on photometry at long wavelength. This is a range where the Galaxy is still quite bright and therefore where background estimation and subtraction becomes critical. As can be seen, the background in the vicinity of the Crab is complex, asymmetric with respect to the object, such that determining it in an aperture around the target can lead to errors or underestimation of the uncertainties. Our approach attempts at capturing all the uncertainties involved in the photometry in presence of complex structural perturbations.}\\

We thus draw the attention to an often overlooked aspect of the difficulty in measuring dust masses in resolved objects, and that is the construction of the spectral energy distribution of the object. This is even more complex in regions where the environment (background or foreground) of the object contributes emission at the location of the object. In this paper we have taken special care to the photometric approach of separating morphologically the background from the Crab nebula. Our proposed technique is a fully objective one which utilizes the gradient variation for choosing the optimum ROI and enabled to construct an objective SED. This is likely safer, and at least more reproducible, than techniques that rely on apertures, however carefully chosen, that appear to match the object, and it allows the adoption of a data driven-approach for the mass and temperature inference. We consider further that, after having applied this method {\em independently} on 6 different photometric images resulting from 2 different instrument (PACS and SPIRE), i.e. a data set where the point spread function, spatial sampling, and noise properties change quite significantly, our method arrives at geometric areas that are very close to each other in each band, as we would expect if the method always extract the same object. Thus even though our dust excess SED is quite different at long wavelengths from previous derivation using similar data, we believe the method used to extract it from the data is more robust, and therefore the derivation of dust mass less dependent on measurement choices. 

We also, remark that the dust mass range obtained by \citet{tea2013}, between 0.02 and 0.13 \Mo, using very different modeling assumption, also brackets our result. {\bf We must however point that the photometric uncertainties that we have established in this paper implies that determining which dust model best fits the observation remains essentially unconstrained. More than sensitivity, improving in this situation will require a better understanding of how to separate the object from its background, which can likely be done by a combination of (a) an better statistical study of the background properties, and (b) higher spatial resolution mapping in the longest wavelengths.} \\




\pagebreak
\section*{Appendix A: Results of calculation}
\label{appendix}
\begin{figure*}[htp!]
\center
	
   {\subfloat{
   \hspace*{-1.5cm}
      \includegraphics[width=0.55\textwidth]{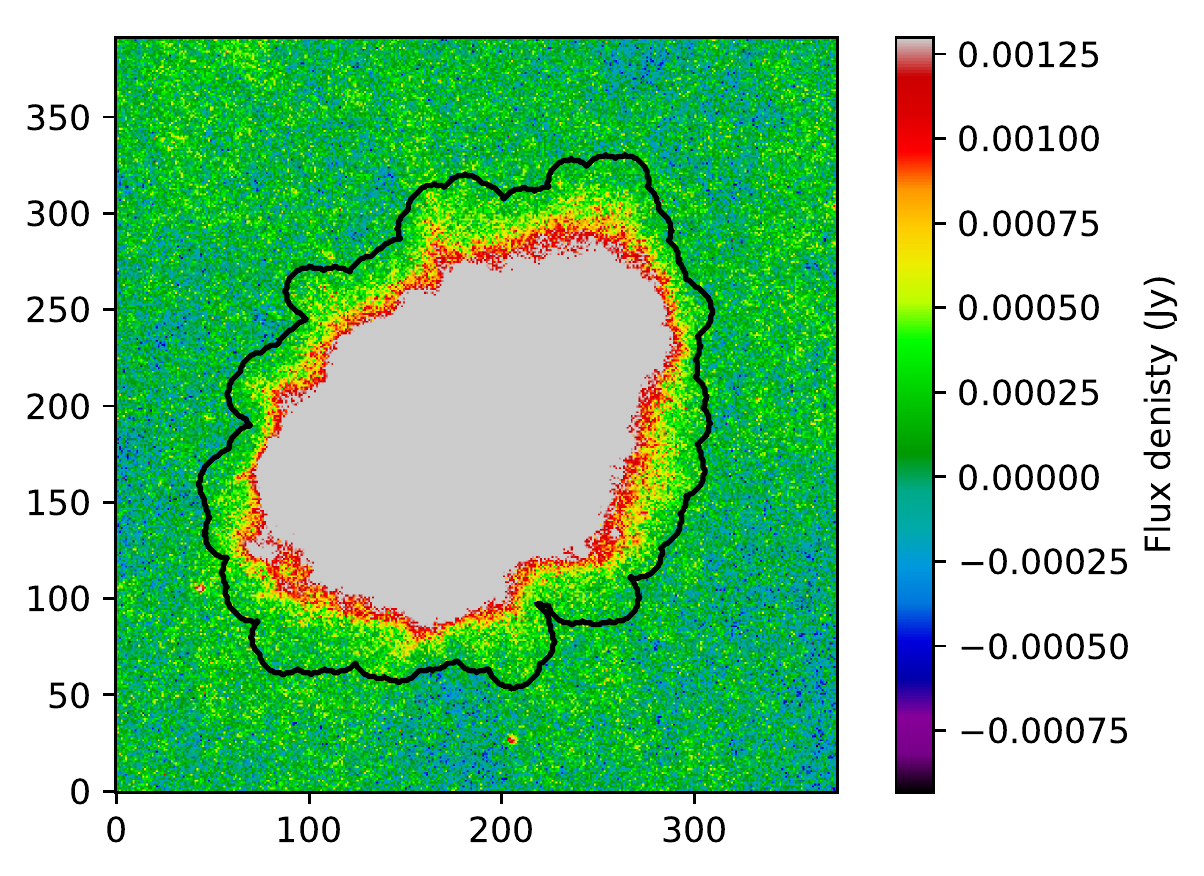}}
~
 \subfloat{
 
      \includegraphics[width=0.55\textwidth]{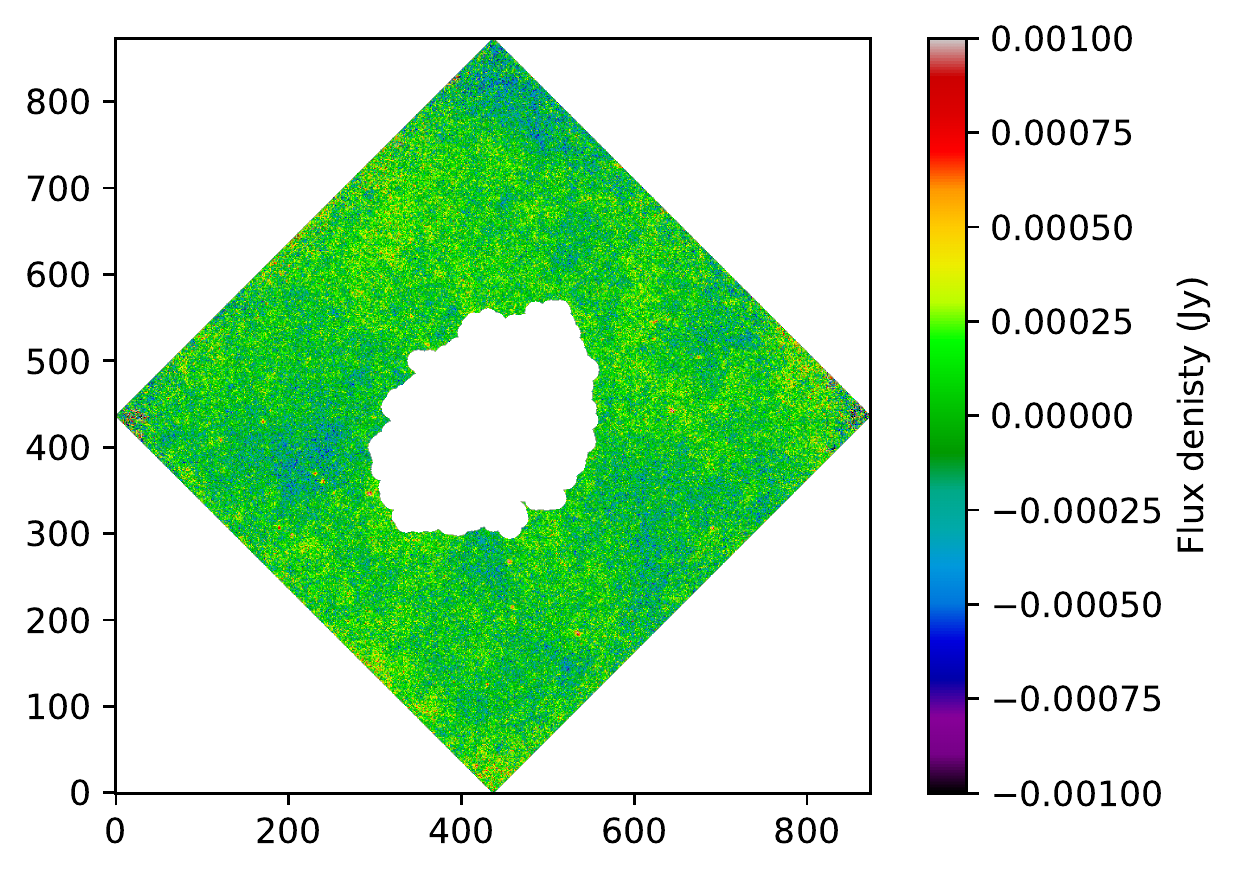}}

}

\vspace{-0.7cm}
\center
   {\subfloat{
\hspace*{-1.5cm}   
      \includegraphics[width=0.55\textwidth]{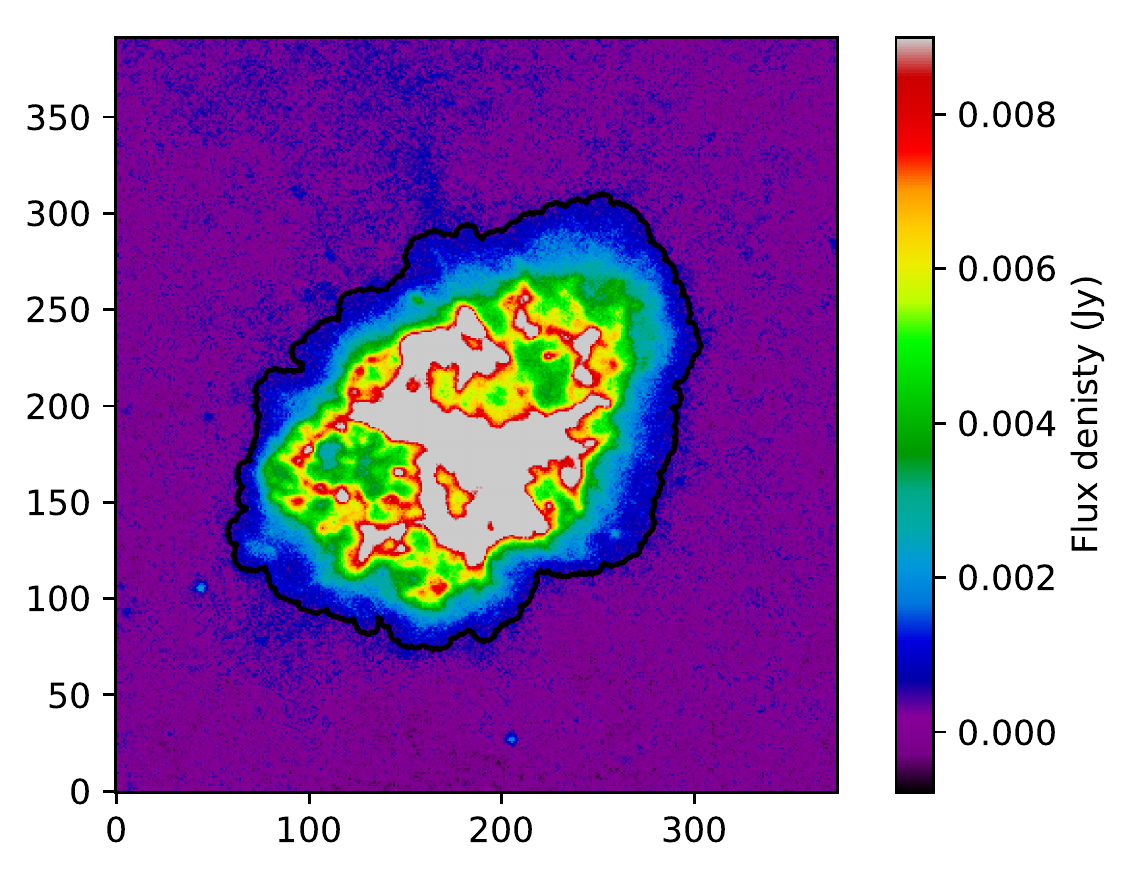}}
~
   \subfloat{
      \includegraphics[width=0.55\textwidth]{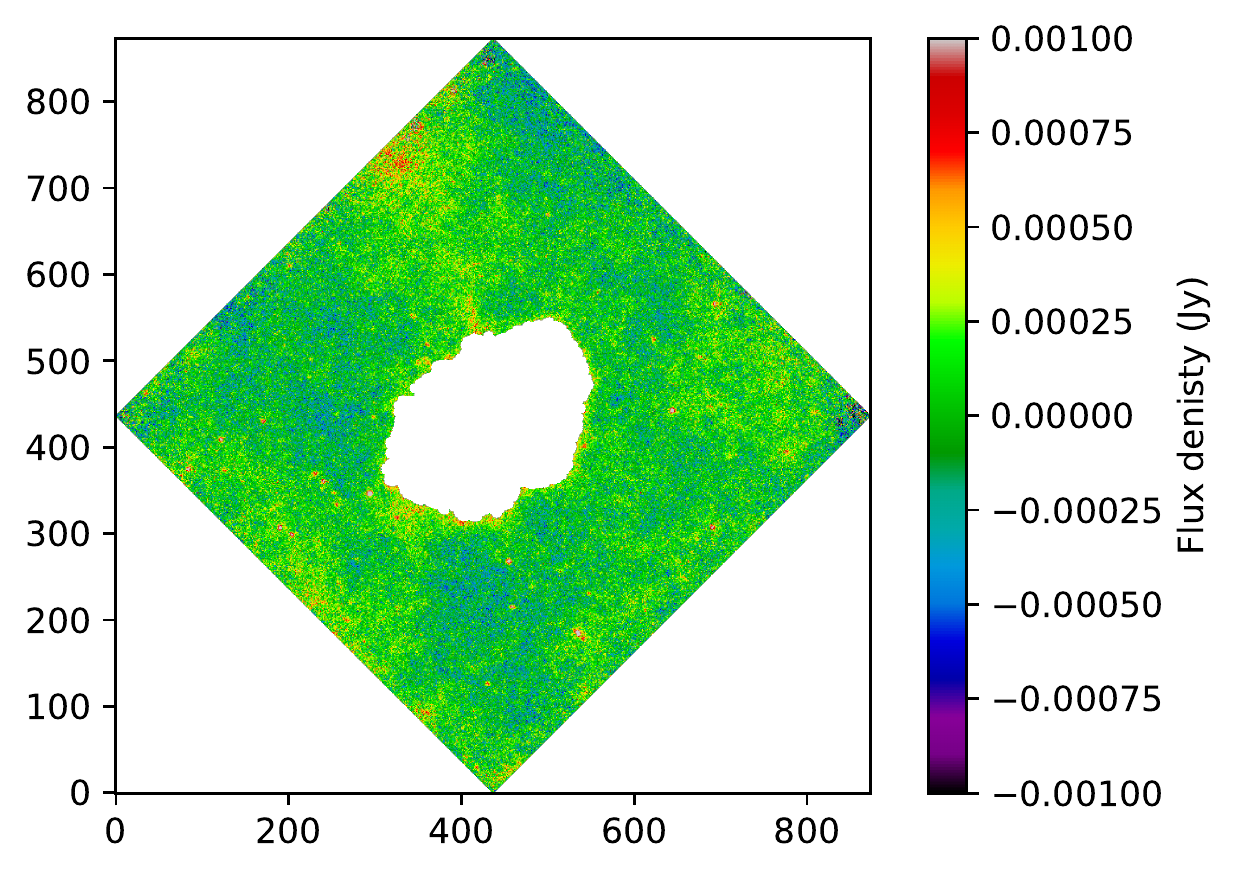}}
   }

 \vspace*{-0.6cm}
    \center
	   
   {\subfloat{
 	  \hspace*{-1.3cm}
      \includegraphics[width=0.55\textwidth]{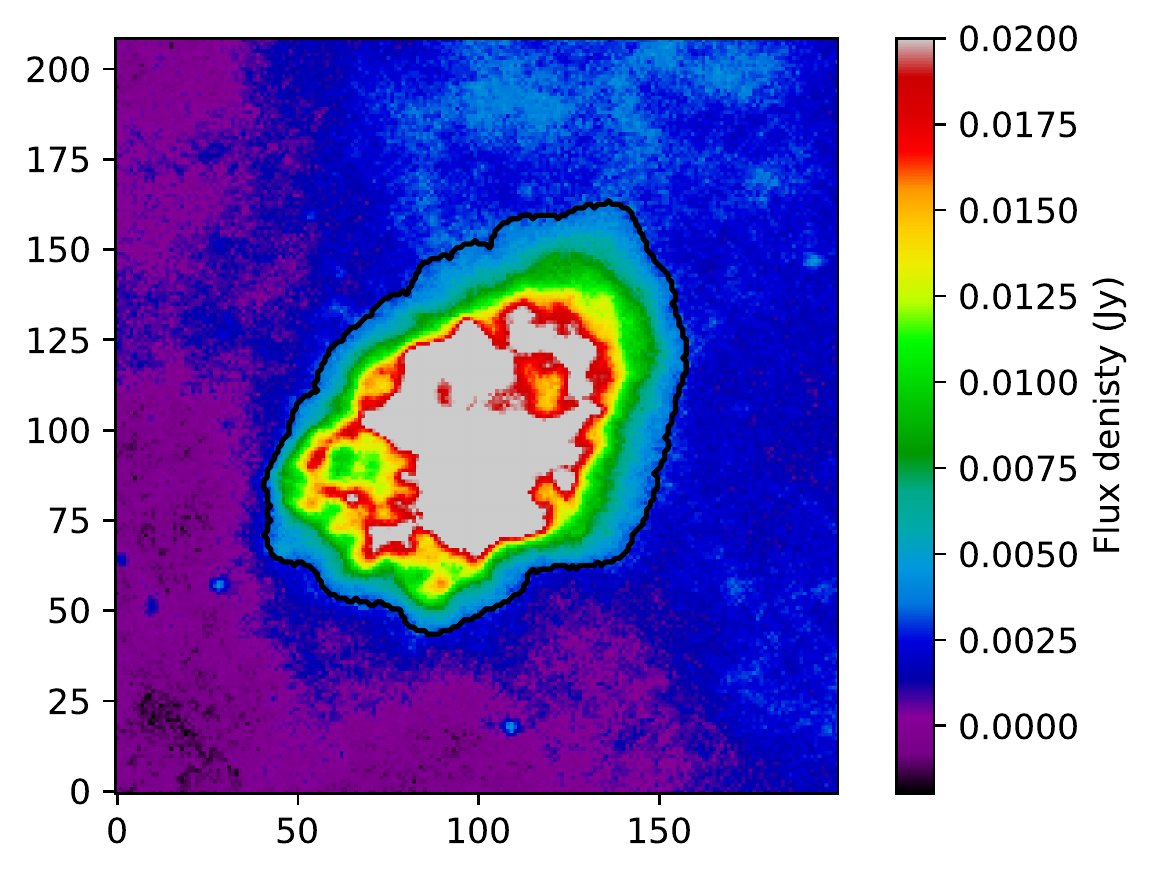}}
~
   \subfloat{
    \hspace*{-0.2cm}
      \includegraphics[width=0.55\textwidth]{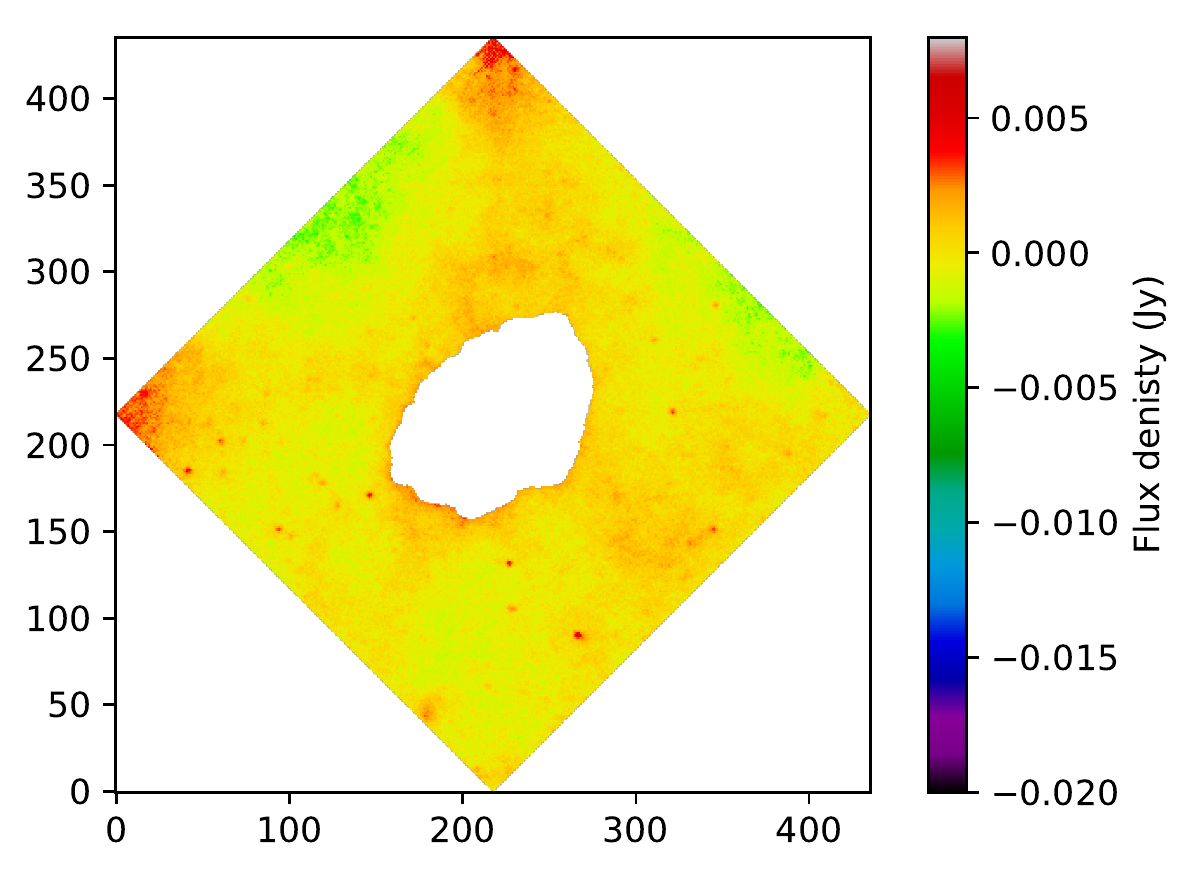}}
   }

   \caption{ Maps of PACS } \label{fig:PACS}
\end{figure*}

\begin{figure*}[htp]
\center
	
   {\subfloat{
   \hspace*{-1.5cm}
      \includegraphics[width=0.55\textwidth]{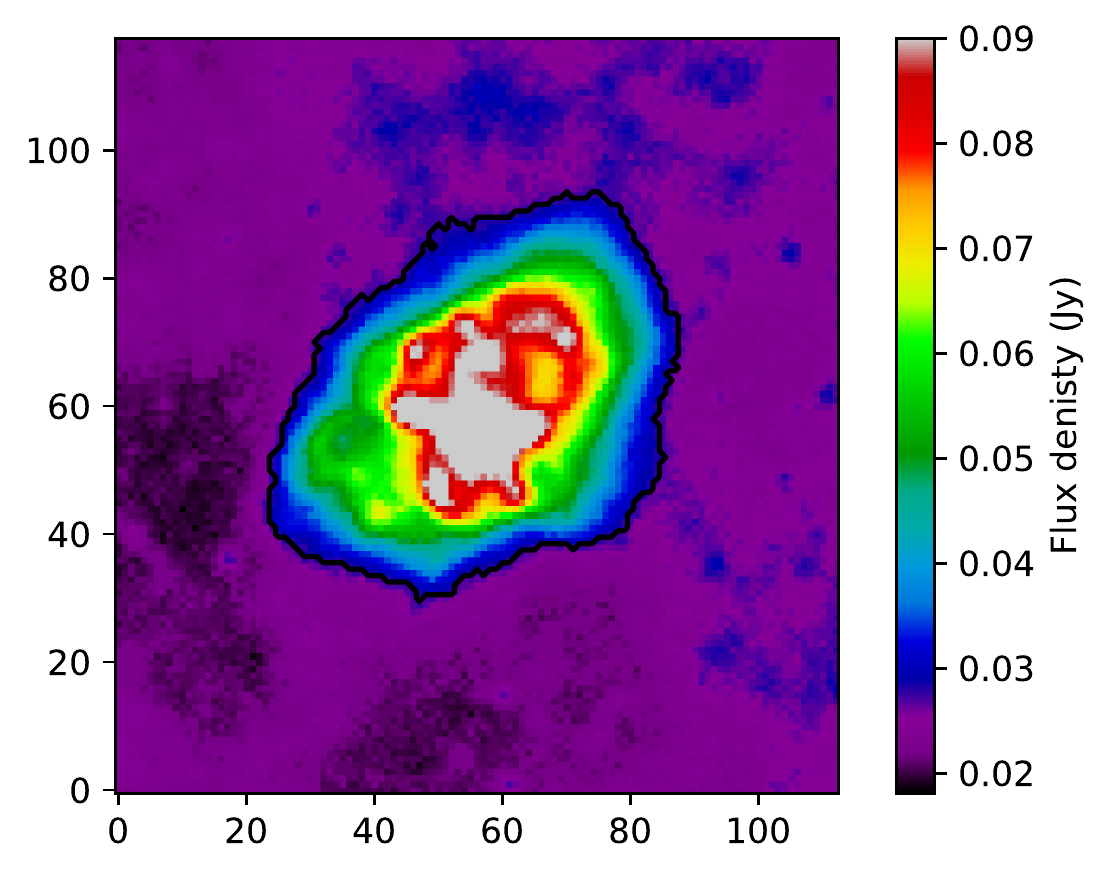}}
~
 \subfloat{
 
      \includegraphics[width=0.55\textwidth]{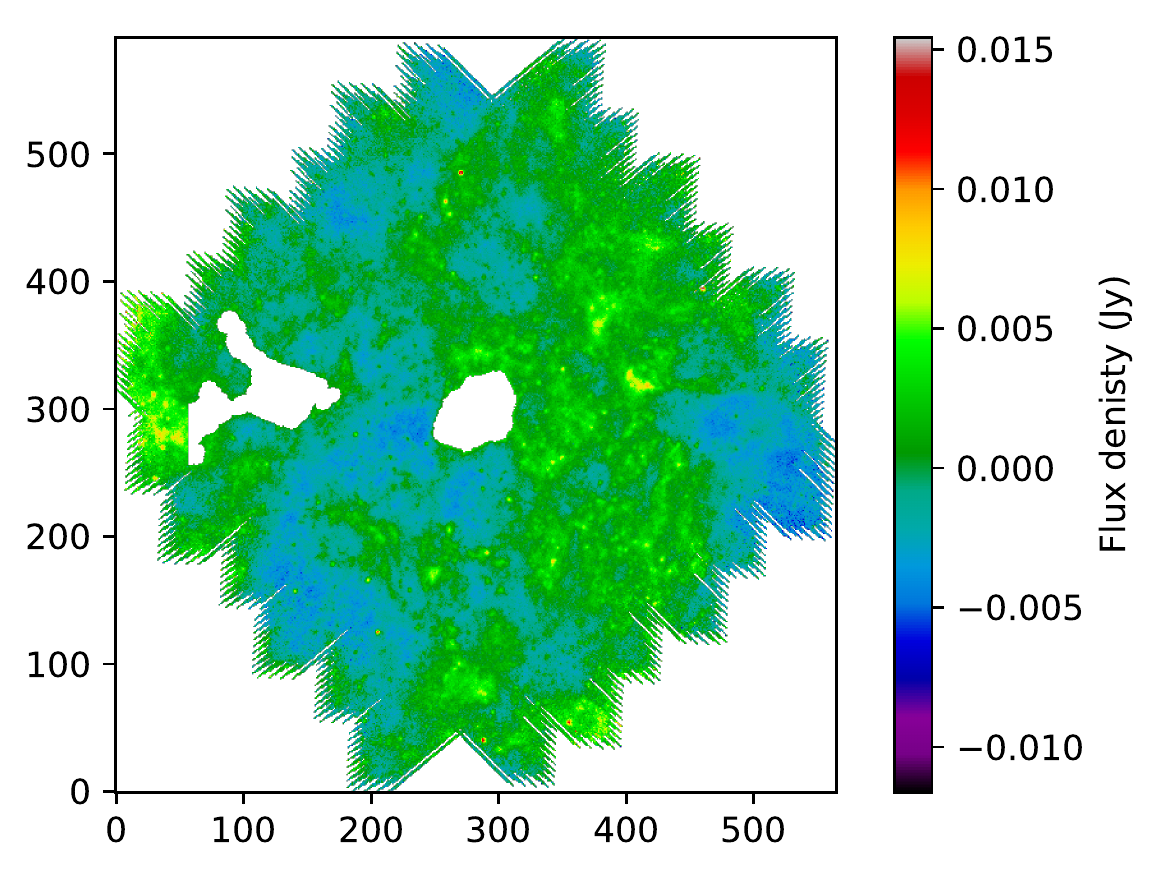}}

}

\vspace{-0.5cm}
\center

 \vspace*{-0.6cm}
    \center
	   
   {\subfloat{
 	  \hspace*{-1.3cm}
      \includegraphics[width=0.55\textwidth]{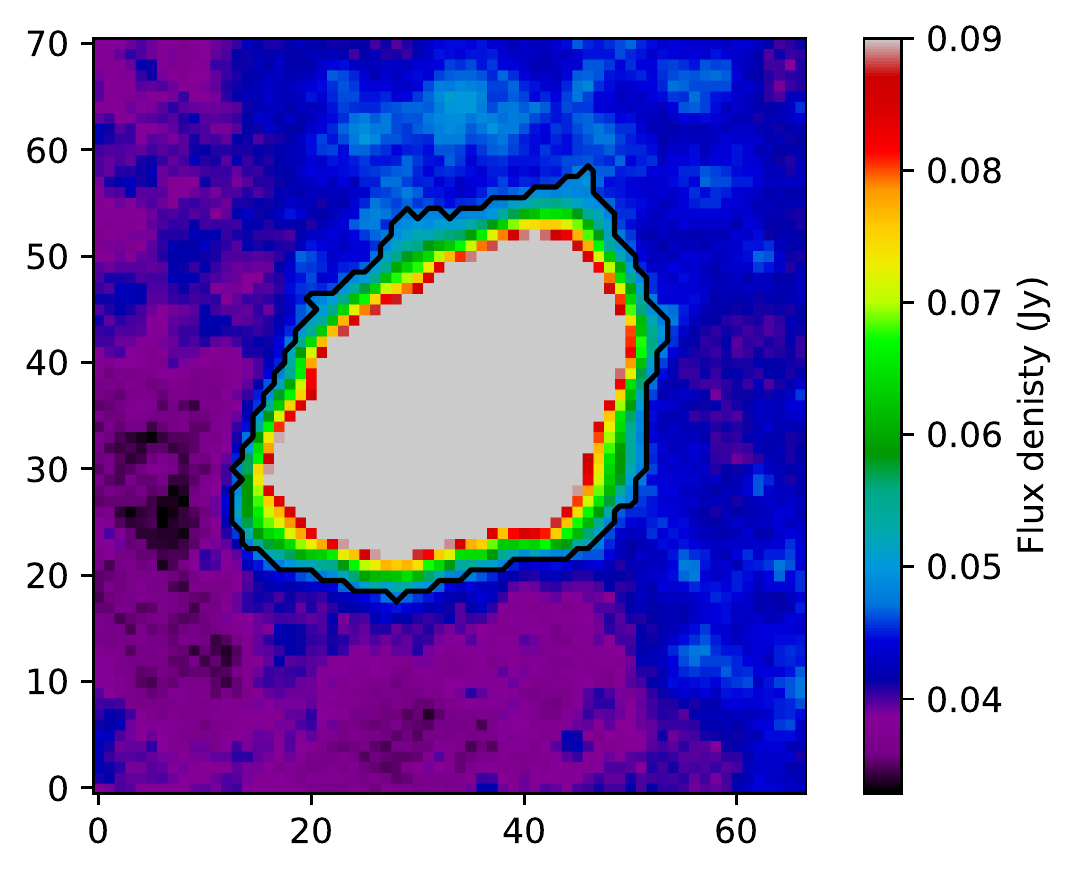}}
~
   \subfloat{
   
      \includegraphics[width=0.55\textwidth]{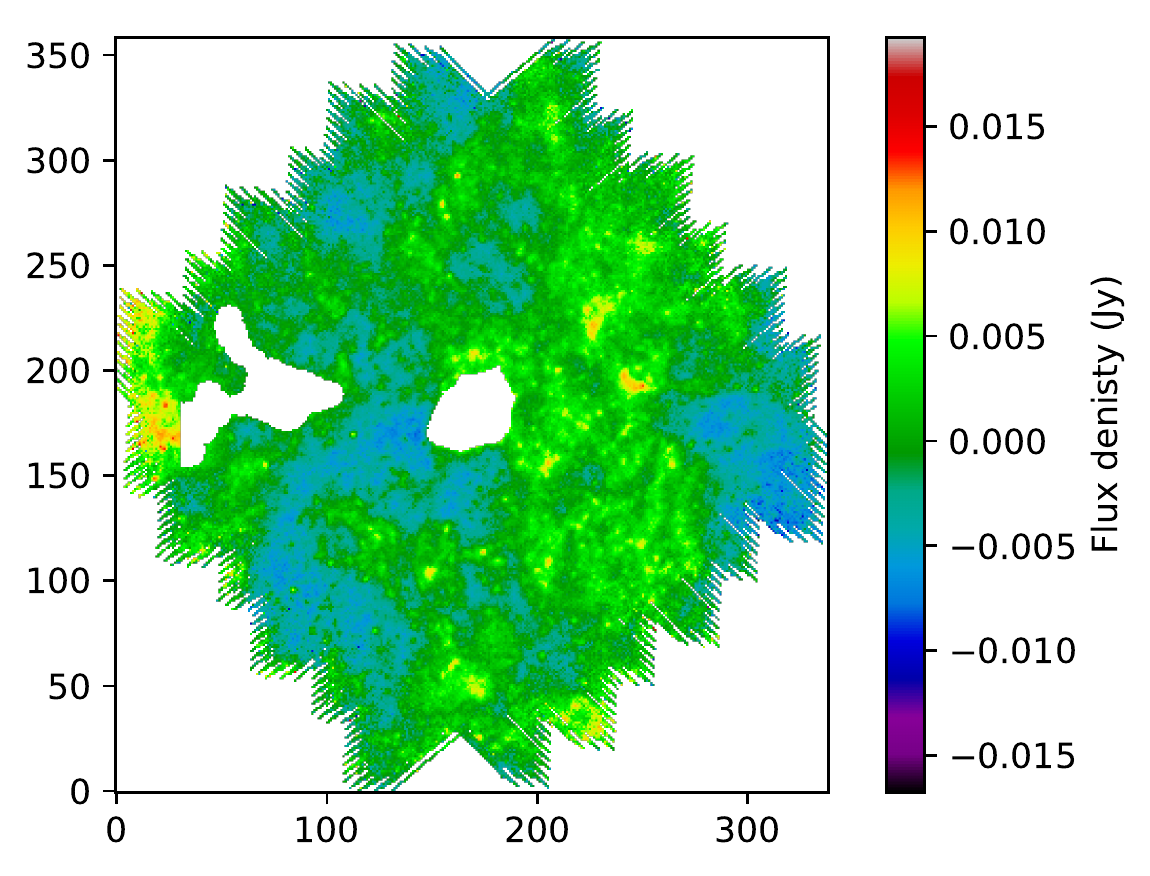}}
   }  
   
   \vspace*{-0.5cm}

    {\subfloat{
 	 \hspace*{-1.3cm}
      \includegraphics[width=0.55\textwidth]{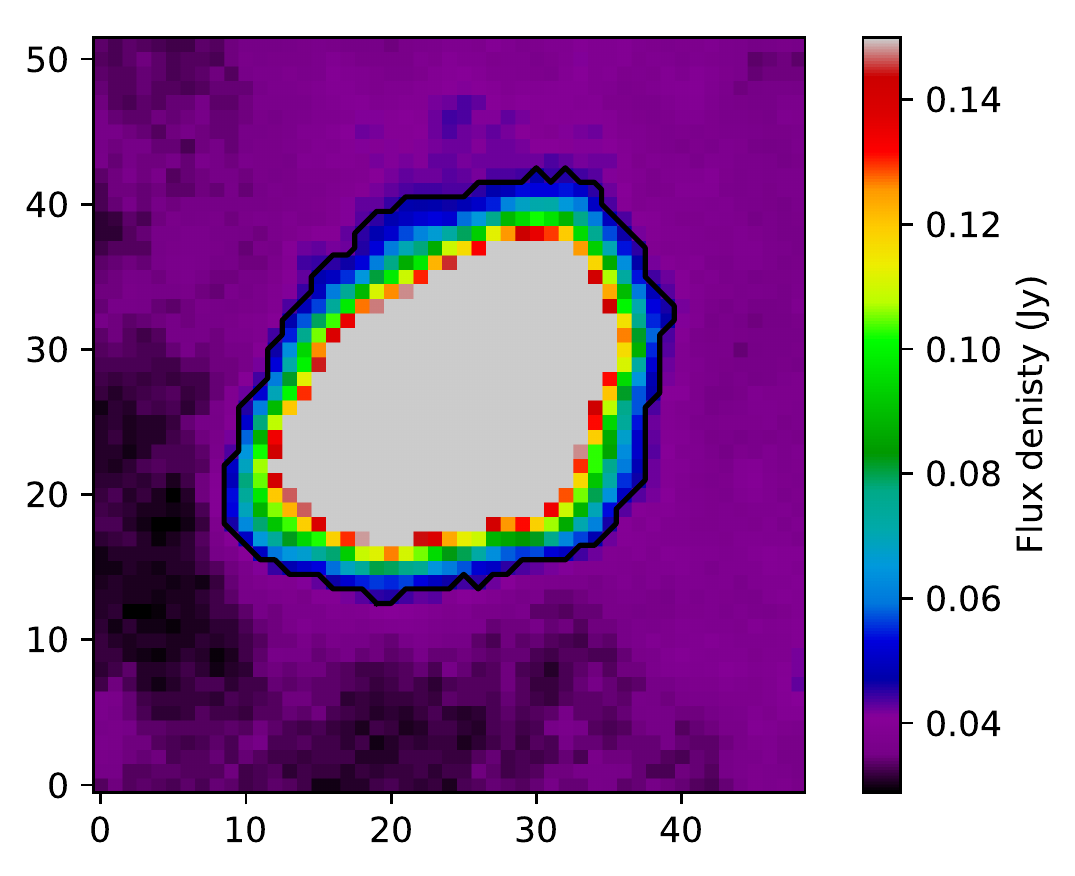}}
~
   \subfloat{

      \includegraphics[width=0.55\textwidth]{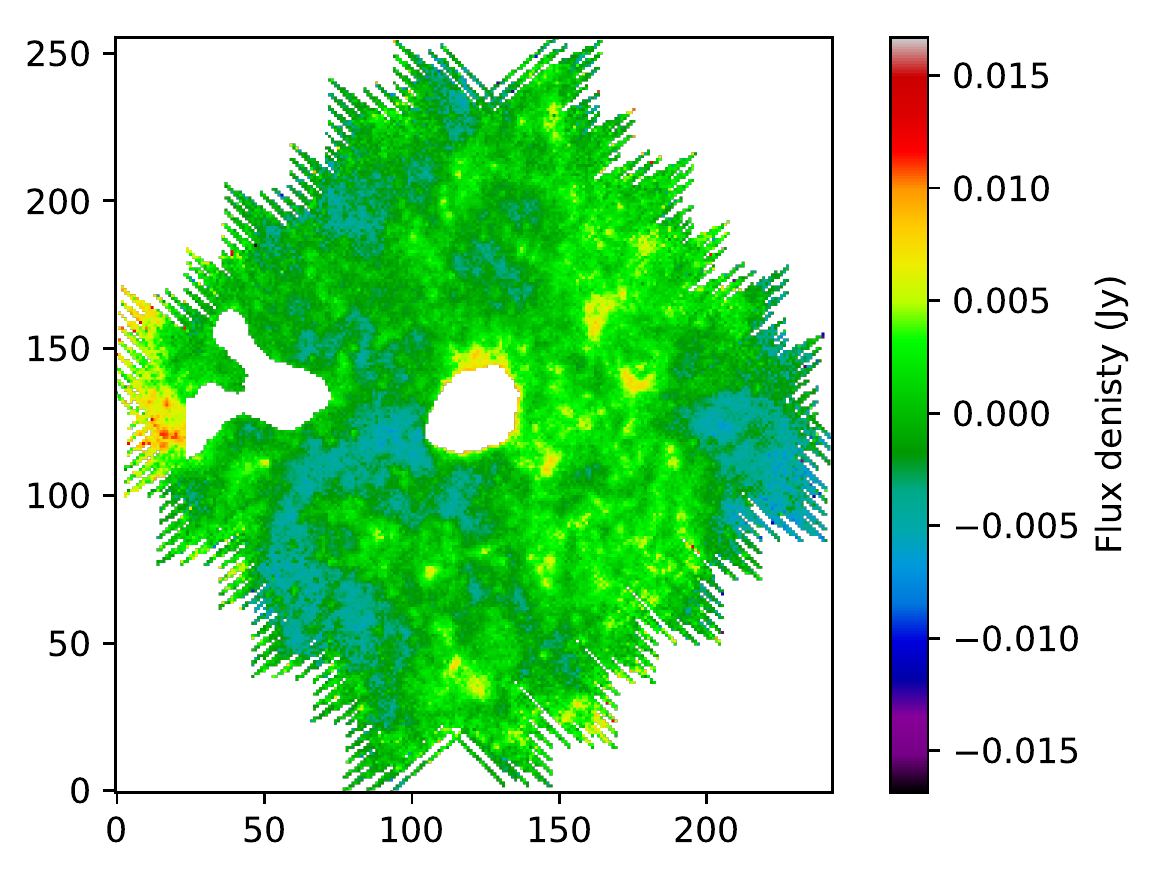}}
   }

   \caption{ Maps of SPIRE } \label{fig:SPIRE}
\end{figure*}

In this appendix, we present a set of visual support material for the results of the image processing of the {\em Herschel} maps. In Fig.~\ref{fig:PACS}, we present the results of the PACS images: For each first,second, and third row corresponding to 70, 100, and 160 \micron\, respectively, the right column  represents background subtracted region of FOV (row to a 1500") and the left column is the zoomed-in region of the crab nebula with the optimum ROI. In Fig.~\ref{fig:SPIRE}, we present the results of the SPIRE images: For each first,second, and third row corresponding to 250, 350, and 500 \micron\, respectively, the right column represents background subtracted region of the full FOV (row to a 1500")and the left column is the zoomed-in region of the crab nebula with the optimum ROI.

\end{document}